\documentclass[usenatbib]{mnras}

\usepackage{newtxtext,newtxmath}

\usepackage[T1]{fontenc}
\usepackage{caption}
\usepackage{subcaption}

\DeclareRobustCommand{\VAN}[3]{#2}
\let\VANthebibliography\thebibliography
\def\thebibliography{\DeclareRobustCommand{\VAN}[3]{##3}\VANthebibliography}

\usepackage{graphicx}	
\usepackage{amsmath}	
\usepackage{lipsum}
\usepackage{float} 
\usepackage{layouts}

\makeatletter
\newcommand\thefontsize[1]{{#1 The current font size is: \f@size pt\par}}
\makeatother

\title[A shark in the waters of AT2017bcc]{Time-varying double-peaked emission lines following the sudden ignition of the dormant galactic nucleus AT2017bcc}

\author[E. J. Ridley et al.]{\noindent
E. J. Ridley,$^{1,2}$\thanks{E-mail: eridley@star.sr.bham.ac.uk} M. Nicholl,$^{3}$ C. A. Ward,$^4$ P. K. Blanchard,$^5$ R. Chornock,$^6$ M. Fraser,$^7$ S. Gomez,$^8$
\newauthor S. Mattila,$^{9,10}$ S. R. Oates,$^{1,2}$ G. Pratten,$^{1,2}$ J. C. Runnoe,$^{11}$ P. Schmidt,$^{1,2}$ K. D. Alexander,$^{12}$ M. Gromadzki,$^{13}$
\newauthor A. Lawrence,$^{14}$ T. M. Reynolds,$^{9,15,16}$ K. W. Smith,$^3$  \L. Wyrzykowski,$^{13}$ A. Aamer,$^{2,3}$ J. P. Anderson,$^{17,18}$
\newauthor S. Benetti,$^{19}$ E. Berger,$^{20}$ T. de Boer,$^{21}$ K. C. Chambers,$^{21}$ T.-W. Chen,$^{22}$ H. Gao,$^{21}$ C. P. Guti\'errez,$^{23,24}$
\newauthor C. Inserra,$^{25}$ T. Kangas,$^{26,27}$ G. Leloudas,$^{28}$ E. A. Magnier,$^{21}$ L. Makrygianni,$^{29}$ T. Moore,$^{3}$
\newauthor T. E. M{\"u}ller-Bravo,$^{23,24}$ S. J. Smartt,$^{30,3}$ K. V. Sokolovsky,$^{31,32}$ R. Wainscoat,$^{21}$ D. R. Young$^{3}$
\\\\
Affiliations are listed at the end of the paper
}

\date{Accepted XXX. Received YYY; in original form ZZZ}

\pubyear{2022}

\begin{document}
\label{firstpage}
\pagerange{\pageref{firstpage}--\pageref{lastpage}}
\maketitle

\begin{abstract}
We present a pan-chromatic study of AT2017bcc, a nuclear transient that was discovered in 2017 within the skymap of a reported burst-like gravitational wave candidate, G274296. It was initially classified as a superluminous supernova, and then reclassified as a candidate tidal disruption event. Its optical light curve has since shown ongoing variability with a structure function consistent with that of an active galactic nucleus, however earlier data shows no variability for at least 10 years prior to the outburst in 2017. The spectrum shows complex profiles in the broad Balmer lines: a central component with a broad blue wing, and a boxy component with time-variable blue and red shoulders. The H$\alpha$ emission profile is well modelled using a circular accretion disc component, and a blue-shifted double Gaussian which may indicate a partially obscured outflow. Weak narrow lines, together with the previously flat light curve, suggest that this object represents a dormant galactic nucleus which has recently been re-activated. Our time-series modelling of the Balmer lines suggests that this is connected to a disturbance in the disc morphology, and we speculate this could involve a sudden violent event such as a tidal disruption event involving the central supermassive black hole, though this cannot be confirmed, and given an estimated black hole mass of $\gtrsim10^7-10^8$\,M$_\odot$ instabilities in an existing disc may be more likely. Although we find that the redshifts of AT2017bcc ($z=0.13$) and G274296 ($z>0.42$) are inconsistent, this event adds to the growing diversity of both nuclear transients and multi-messenger contaminants.
\end{abstract}


\section{Introduction}
It is now established that most galaxies host a central supermassive black hole (SMBH) \citep{Magorrian1998,Kormendy2013}. Accretion onto the SMBH can allow a galactic nucleus to outshine its host. Flickering like candles, these active galactic nuclei (AGN) vary in luminosity over time typically at the level of a few tenths of a magnitude as the rate of their accretion changes. This emission illuminates gas hundreds of light years from the SMBH, producing narrow-line spectral features. In some AGN, Doppler-broadened emission is also visible, arising from fast-moving gas surrounding the central accretion disc. In the optical, this is mainly visible in the Balmer lines. While almost all AGN show variable luminosity and narrow-line emission, they are categorised based on whether the broad-line region is visible. Type I AGN show both broad- and narrow-line features, while Type II AGN lack the broad-line component. This is attributed to a viewing-angle effect, with the broad line region obscured by a dusty torus in Type II AGN \citep{Antonucci1993,Urry1995}.

Due to their persistent variable emission, AGN have historically been treated as contaminants during searches for transient objects. However, with the rise of wide-field sky surveys and automated source detection, we have been able to uncover the more unusual behaviour of nuclear emitters. For example, some AGN have been seen to transition from Type I and Type II, or vice versa, between epochs. These are known as changing-look or changing-state AGN and such objects challenge the paradigm that the difference between the types is purely viewing angle. Though we have observations dating back to the 1970s \citep{Khachikian1971,Antonucci1983}, they have been detected more frequently in the last decade \citep{LaMassa2014,Runnoe2016,Lawrence2016,MacLeod2019,Ricci2022}.

We have also learned that AGN are not the only source of luminous emission from galactic nuclei. All SMBHs can produce flares by tidally disrupting stars that wander too close, even those which are non-accreting or "quiescent". The existence of such tidal disruption events (TDEs) was proposed in the 1970s and 80s \citep{Hills1975,Rees1988}, and the first candidates were identified in the late 1990s \citep{Komossa1999}. Since then, we have detected almost $100$ TDEs and broadly classified them based on whether they radiate in the X-ray \citep{Auchettl2017,Sazonov2021}, optical \citep{Gezari2012,Arcavi2014,Holoien2014,vanVelzen2021}, or both and if they produce a relativistic jet \citep{Alexander2020,Andreoni2022, Cendes2022, Pasham2023}. As with AGN, some of this diversity of emission is thought to be a result of viewing angle dependence, arising from a non-spherically-symmetric envelope of reprocessing material around the SMBH \citep{Dai2018}.

AGN accretion discs may also contain sources of gravitational wave (GW) emission. It is expected that galactic nuclei host a dense population of stellar-mass black holes, many of which reside in the plane of the central accretion disc \citep{Morris1993}. Binary black holes (BBHs) in such an environment would merge rapidly due to drag from the surrounding gas \citep{McKernan2020}. Shocks in the gas and super-Eddington accretion onto the black holes would also produce a fast, bright electromagnetic (EM) transient \citep{Bartos2017,McKernan2019}. Thus, these mergers could be multi-messenger events, producing both GW and EM emission. A candidate for such an event was reported by \cite{Graham2020}, along with the proposal to monitor AGN when searching for future EM counterparts to GW -- including BBH -- detections.

In this paper, we present observations of a nuclear transient, AT2017bcc, which was first observed as a candidate EM counterpart to a GW detection. We explore the possibility that this is a genuine multi-messenger event by re-analysing the GW signal, finding that the GW and EM sources are likely unrelated. However, the EM source itself shows a number of unusual properties that provide insight into the diversity of AGN variability and nuclear transients, and may shed light on the nature of the AGN changing look phenomenon. We have obtained an extensive data set from X-ray to radio, including time series spectroscopy over a period of six years.

In particular, this event is a rare example (unique to our knowledge) of an AGN that is both changing state and a `double-peaked emitter' \citep{StorchiBergmann1993,Eracleous1994} with complex broad-line profiles that vary over time \citep{Gezari2007}. Double-peaked emitters often exhibit distinct blue and red shoulders in their broad Balmer lines, consistent with an accretion disc origin \citep{Eracleous1995}. Understanding the line profiles and their evolution in AT2017bcc provide new clues to the processes that have switched this AGN on after a period of quiescence.

The paper is structured as follows. In Section 2 we describe the discovery of AT2017bcc and our re-analysis of G274296. We detail our spectroscopic and photometric follow-up in Section 3. Using fits to archival photometry, we examine the host galaxy in Section 4. We interpret the multi-wavelength light curve in Section 5. In Section 6 we present our optical spectroscopy and model the emission profiles to infer physical parameters. Finally we discuss our results in Section 7, and conclude in Section 8. Unless otherwise stated, we adopt a flat $\Lambda$CDM cosmology with $H_{\rm 0} = 67.7\,{\rm kms^{-1}}$, $\Omega_{\rm 0} = 0.31$ throughout \citep{Planck2020}.

\section{Discovery}
On $17^{\rm th}$ February 2017, the LIGO collaboration reported the identification of a candidate signal, labelled G274296 \citep{2017GCN.20689....1L}, during real-time burst analysis. The signal was flagged by the \textsc{coherent waveBurst} (CWB; \citealt{Drago2020}) pipeline; designed to identify GW transients in detector streams without prior knowledge of a signal waveform. As such, G274296 did not obviously resemble the typical waveform "chirp" expected from a binary inspiral (though this could not be ruled out), but was still significant due to its low false alarm rate of $\sim1$ per 2 months.

During the subsequent search for EM counterparts, the Panoramic Survey Telescope and Rapid Response System (Pan-STARRS; \citealt{Chambers2016}) discovered a bright nuclear transient on $19^{\rm th}$ February, then dubbed PS17bgn, in the galaxy SDSS J113152.97+295944.8 within the 90\% contour of the G274296 skymap \citep{Chambers2017}. It was spectroscopically classified by the Public ESO Spectroscopic Survey of Transient Objects (PESSTO; \citealt{Smartt2015}) as a type II superluminous supernova (SLSN-II) due to its broad H$\alpha$ emission at redshift $z=0.133$, and corresponding absolute magnitude of $-21$ \citep{2017GCN.20708....1T}. The transient was then renamed to SN2017bcc and dropped from the counterpart search. Later re-examination by \cite{2017ATel10177....1S} revealed blue optical and near-UV continuum, suggesting high temperatures. They also detected non-thermal X-ray emission. These features are characteristic of tidal disruption events (TDEs) and active galactic nucleus (AGN) flares, and though they do not rule out a SLSN-II origin, X-rays have not been detected in the majority of such objects.

Given the uncertainty in the classification of this transient (and the results of our own analysis), we will refer to this source using its pre-classification IAU name AT2017bcc.

\subsection{GW analysis}
\begin{figure}
    \centering
    \includegraphics[width=\linewidth, trim=0cm 1cm 0.5cm 0.5cm]{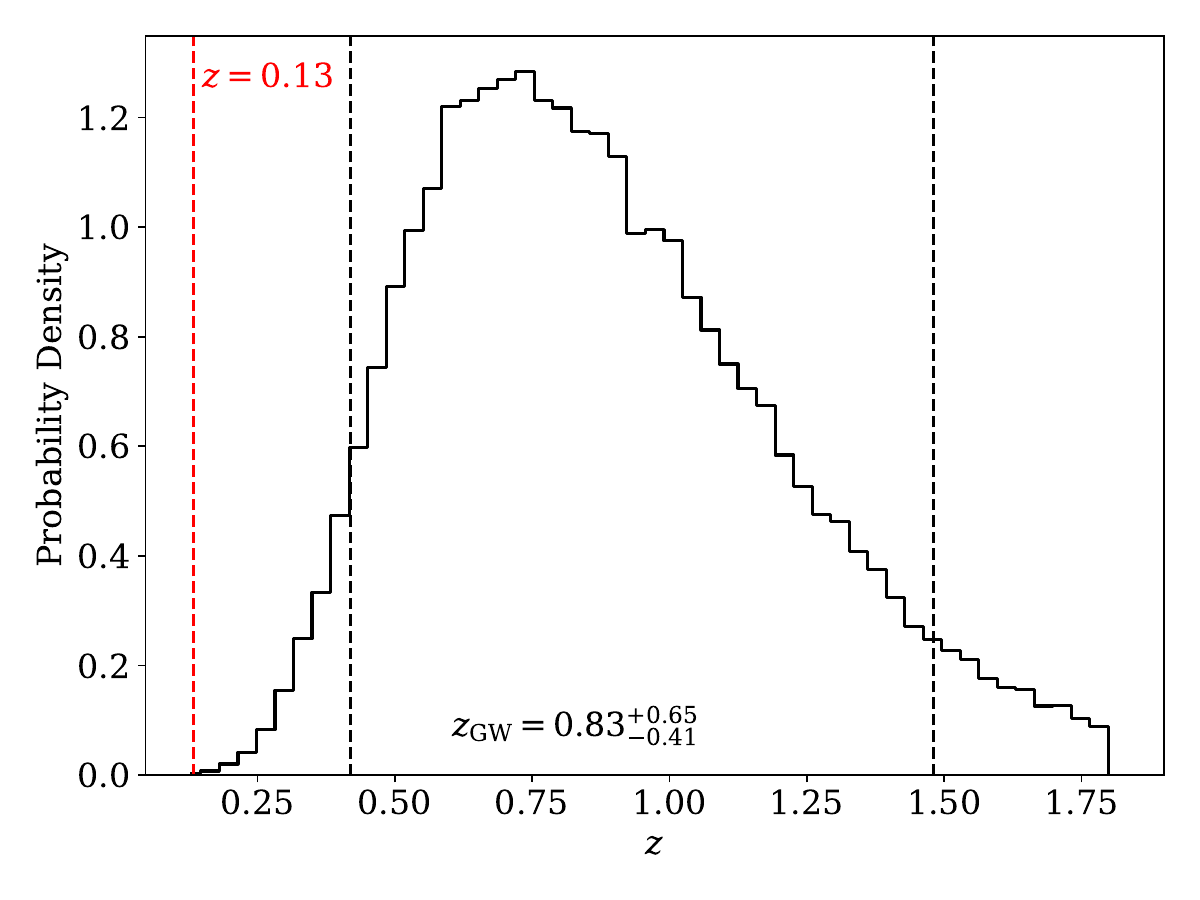}
    \caption{Posterior probability density on redshift from waveform analysis of G274296. The vertical dashed lines indicate the $1\sigma$ region (black) and the spectral redshift of AT2017bcc (red).}
    \label{fig:gw_z}
\end{figure}

\begin{figure*}
    \centering
    \includegraphics[width=\linewidth, trim=0cm 1cm 0cm 0cm]{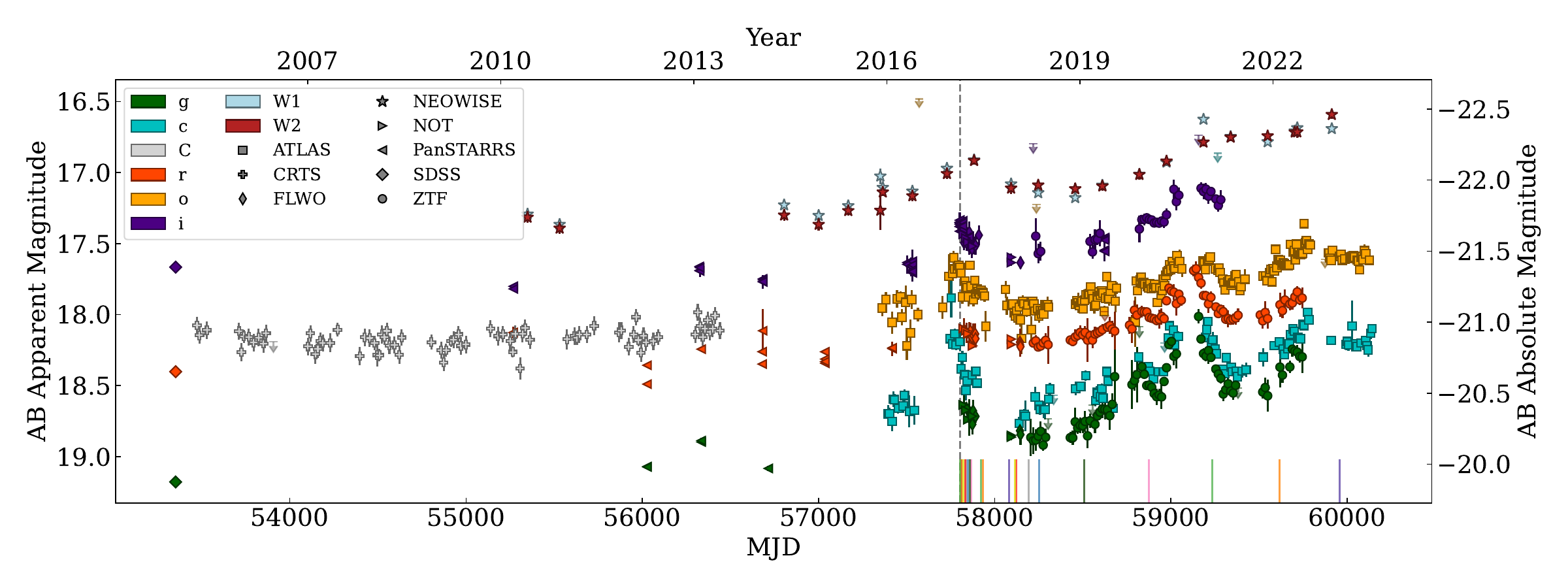}
    \caption{Optical and infra-red light curve of AT2017bcc. All magnitudes are corrected for galactic extinction and still include host light. Each point is the result of binning fluxes to a weekly cadence for visual clarity. Epochs below a $5\sigma$ detection limit are plotted as upper limits with a lower opacity. The vertical dashed line indicates the discovery by Pan-STARRS in February 2017. The epochs of spectroscopy are marked below the magnitudes, with the same colors used in Figure \ref{fig:all_spec}.}
    \label{fig:at2017bcc_lc}
\end{figure*}

Since AT2017bcc was discovered during optical follow-up of a GW source, we here assess whether the EM transient could be the true counterpart to the GW signal.

As G274296 was identified by a GW pipeline using unmodeled analysis, a distance to the source could not be determined in real time. Burst-style gravitational wave signals are expected to originate from stellar core collapse, thought to be detectable at current GW sensitivity only at kiloparsec-scale distances \citep{Gossan2016} (though collapse to a rapidly rotating black hole -- a collapsar -- may be detectable at $\sim100$\,Mpc; \citealt{Gossan2016}). As such, it is unlikely that G274296 is a true burst source, as it would most likely have been accompanied by an easily identifiable nearby supernova. 
In any case, a GW burst from a stellar core collapse would not be detectable at the distance of AT2017bcc (625 Mpc) \citep{Gossan2016}. Therefore if G274296 did originate in this way, the EM and GW sources are not associated.

The alternative scenario is that the signal arose from a compact binary merger. A sufficiently massive system may enter the LIGO bandpass only in the final stages of merger, such that the distinctive chirp signal from the inspiral phase goes unseen. This is a plausible origin of an AGN flare-like counterpart, if the merger occurred within an accretion disc around an SMBH. Indeed, the only previous EM candidate for a merger in an AGN disc \citep{Graham2020} corresponded to one of the most massive BBH mergers detected by LIGO to date, GW190521: a system with a total mass of $\sim150$\,M$_\odot$ \citep{Abbott2020}. Under the assumption of a compact binary merger, to determine the source properties of G274296 we carried out a full Bayesian analysis~\citep{Veitch:2014wba} using two state-of-the-art waveform models, IMRPhenomXPHM~\citep{Pratten:2020ceb} and NRSur7dq4~\citep{Varma:2019csw}, and found it to be consistent with a very massive compact binary with median source-frame progenitor masses of $119(120)\: \rm{M_\odot}$ and $77(73)\: \rm{M_\odot}$ using IMRPhenomXPHM (NRSur7dq4). In this scenario, due to the large component masses, the GW signal is very short in duration ($\ll 1\: {\rm s}$) and dominated by the merger phase, emulating a burst signal, which is consistent with the initial detection by a burst pipeline.

This analysis also produced a luminosity distance posterior, shown in Figure \ref{fig:gw_z}. In this scenario, using the cosmological parameters from~\citet{Planck:2015fie} we find the median redshift of G274296 to be $z_{\rm GW} = 0.83^{+0.65}_{-0.41}$. The spectroscopic redshift of AT2017bcc, measured from its narrow [\ion{O}{III}] lines, is $z_{\rm EM} = 0.133$ which is outside the 99th percentile of the GW distance posterior. Thus we conclude that, with the absence of further models for the GW signal, AT2017bcc is not a valid EM counterpart for G274296. In the rest of this paper, we will analyse our extensive EM data set and the implications of this very unusual nuclear source for understanding extreme AGN variability.

\section{Observations}
\subsection{Photometry}

\begin{figure}
    \centering
    \includegraphics[width=\linewidth, trim=0cm 1cm 0.5cm 0.5cm]{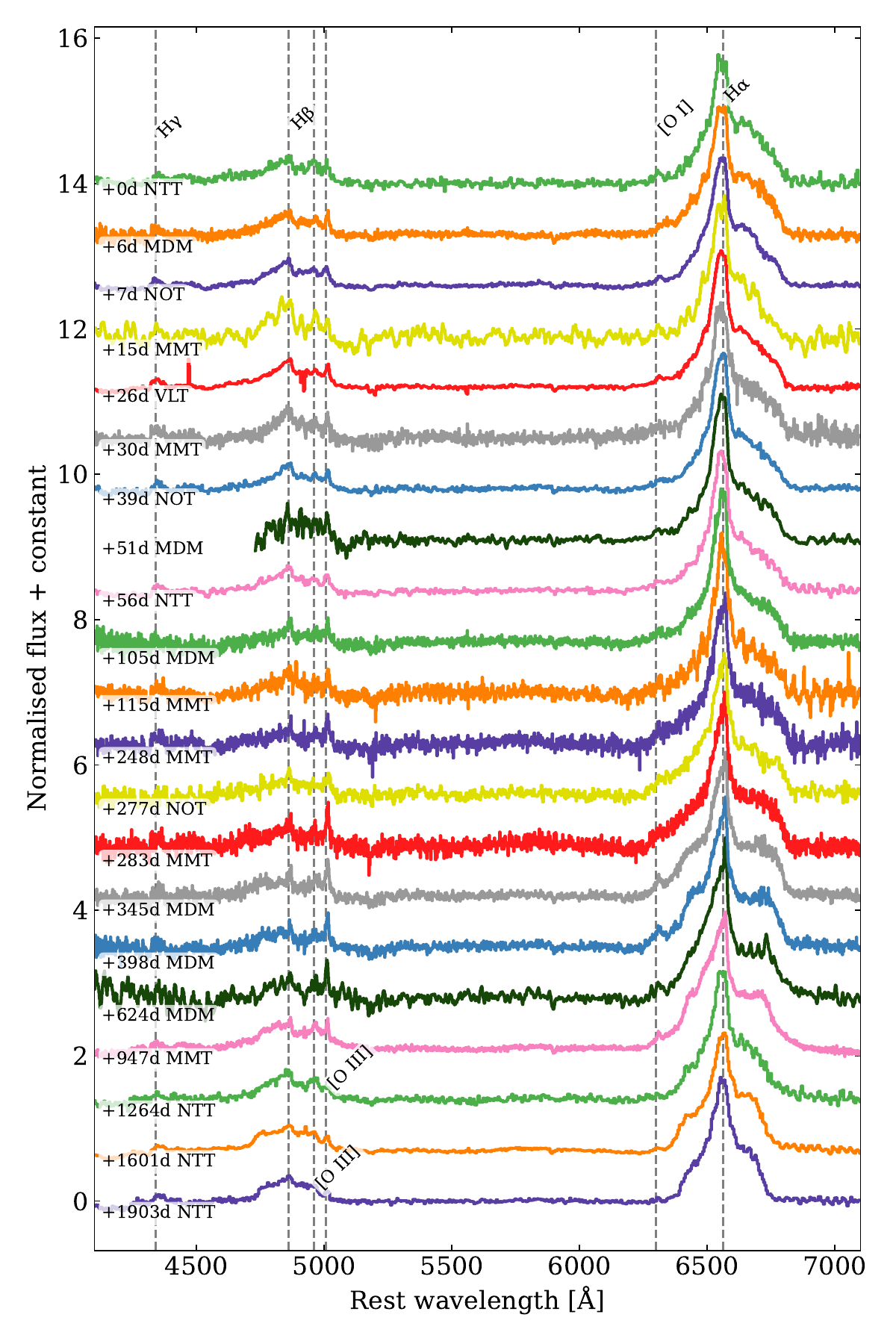}
    \caption{Optical spectroscopic evolution of AT2017bcc, containing 20 epochs from day one ($19^{\rm th}$ February 2017) to day 2157 ($15^{\rm th}$ January 2023) post-discovery. Each epoch is labelled with the rest-frame days since discovery and the telescope / observatory which took the data. All spectra are telluric-corrected, continuum-subtracted, and normalised for visual clarity. Some spectra have been smoothed using a Savitzky-Golay filter.}
    \label{fig:all_spec}
\end{figure}

We observed AT2017bcc during the first optical peak in the \textit{g, r, i} bands using KeplerCam on the 1.2-m telescope at the Fred Lawrence Whipple Observatory (FLWO), and in \textit{g, r, i, z, J, H, K} using the ALFOSC optical imager and spectograph and NOTCam NIR imager and spectrograph on the 2.5-m Nordic Optical Telescope (NOT) at the Roque de los Muchachos Observatory. Bias subtraction and flat fielding were applied either using \textsc{iraf} or instrument specific pipelines, and sky subtraction was applied to the NOTCam images. Aperture photometry with local background subtraction was performed on all images using \textsc{photutils} \citep{Bradley2021} with field star magnitudes from Pan-STARRS and the Two Micron All Sky Survey (2MASS; \citealt{Skrutskie2006}) to calculate the photometric zero-points. SDSS reports a petrosian radius of $2.5$, $2.4$, $1.8$, and $2.6\,\rm{arcsec}$ for the \textit{g, r, i, z} bands respectively. Thus we decided that an aperture size of $5\,\rm{arcsec}$ was sufficient to include flux from the compact host galaxy consistently for our optical bands.

We imaged AT2017bcc in \textit{uvw1, uvw2, uvm2, U, B, V} using the Ultraviolet/Optical Telescope (UVOT; \citealt{Roming2005}) on the Neil Gehrels \textit{Swift} Observatory. Count rates were obtained using the \textit{Swift} \textsc{uvotsource} tool and converted to magnitudes (in the AB system) using the UVOT photometric zero points \citep{Breeveld2011}. The source counts were extracted initially using a source region of $5\,{\rm arcsec}$ radius. When the count rate dropped to below $0.5$ counts per second, we used a source region of $3\,{\rm arcsec}$ radius. In order to be consistent with the UVOT calibration, these count rates were then corrected to $5\,{\rm arcsec}$ using the curve of growth contained in the calibration files. Background counts were extracted using a circular aperture of radius $20\,{\rm arcsec}$ located in a source-free region of the sky.

We also downloaded reduced data from various public surveys covering 2005-2022. Images in \textit{g, r, i, z} were obtained from Pan-STARRS via the PS1 Image Cutout Service; and in \textit{g, r, i} from the Zwicky Transient Facility (ZTF; \citealt{Bellm2019}), via the NASA/IPAC Infrared Science Archive. These were analysed using \textsc{photutils} in order to use a consistent aperture on science images, rather than difference magnitudes, to avoid any unwanted effects from transient contamination in template images.

We acquired survey magnitudes for AT2017bcc in \textit{c, o} from the Asteroid Terrestrial-Impact Last Alert System (ATLAS; \citealt{Tonry2018,Smith2020,Shingles2021}) forced photometry server, using reduced rather than difference images; in \textit{V} from the Catalina Real-time Transient Survey (CRTS; \citealt{Djorgovski2011}) cone search service; in \textit{W1, W2} from the Near-Earth Object Wide-field Infrared Survey Explorer (\textit{NEOWISE}; \citealt{Mainzer2011}) via the NASA/IPAC Infrared Science Archive; and in \textit{u, g, r, i, z} from the Sloan Digital Sky Survey (SDSS; \citealt{Alam2015}).

Figure \ref{fig:at2017bcc_lc} shows all of our optical and infrared (IR) photometry in AB magnitudes. The \textit{Swift} UVOT counts had very large uncertainties so were not included in the light curve for visual clarity, but we do show them in Figure \ref{fig:sed}.

\subsection{X-ray data}\label{sec:xray}

\begin{figure}
    \centering
    \includegraphics[width=\linewidth, trim=0cm 1cm 0.5cm 0cm]{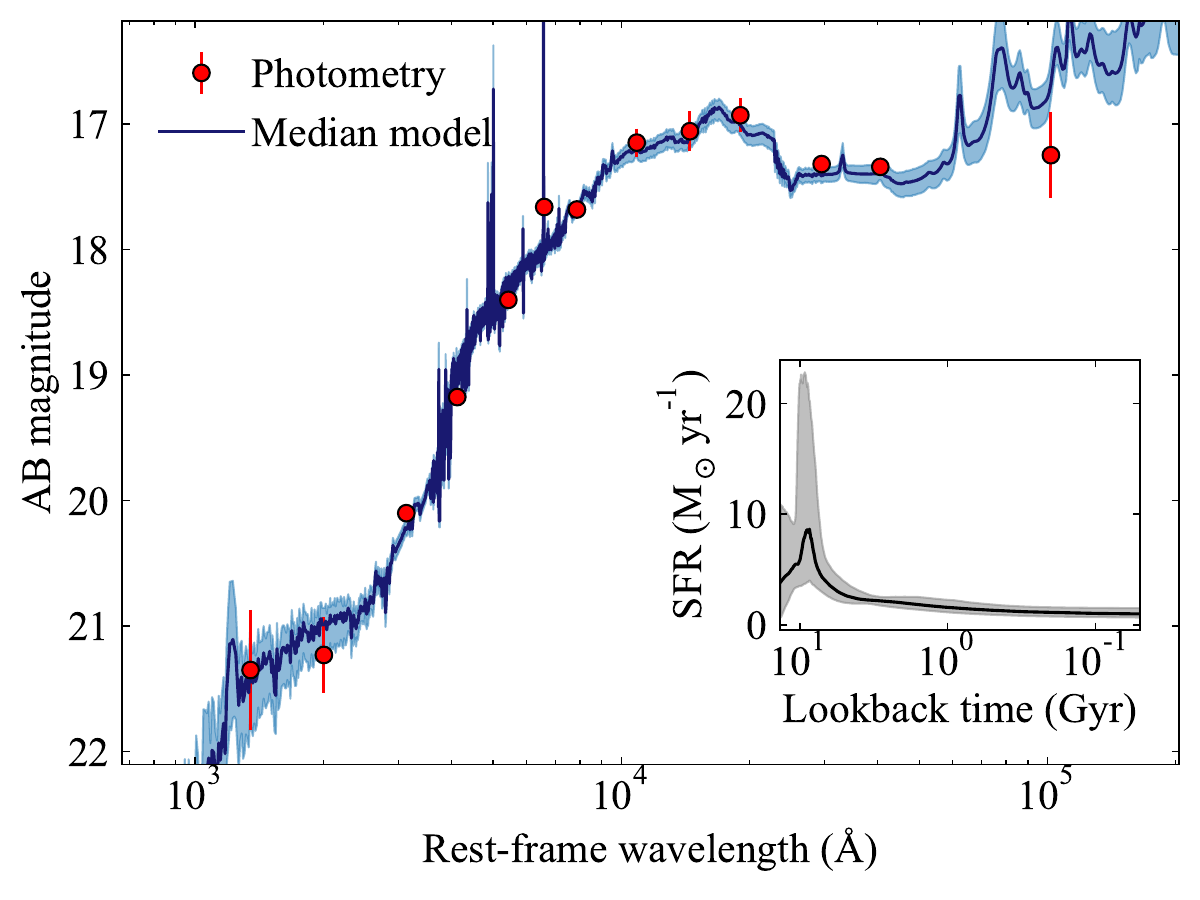}
    \caption{
    Fit to the archival spectral energy distribution of the AT2017bcc host galaxy using \textsc{Prospector}. The inset shows the implied star-formation history. Shaded areas show the 16th and 84th percentiles of the model posteriors.
    }
    \label{fig:at2017bcc_sed_sdss}
\end{figure}

X-ray observations were acquired using the X-ray Telescope (XRT) on board \textit{Swift} over 12 epochs from 2017-03-10 to 2018-03-14. Data were downloaded and analysed using the automated tools provided by the UK Swift Science Data Centre \citep{Evans2007,Evans2009}. We stacked all available data to produce a high-S/N X-ray spectrum from 0.3-10\,keV with a total exposure of 20.5\,ks, and fit this with both power-law and blackbody models at a redshift $z=0.133$. The models include both a redshifted and Galactic absorber, using the T{\"u}bingen-Boulder ISM absorption model \citep[TBabs;][]{Wilms2000} and the Galactic column from \citet{Willingale2013}, fixed at $1.9\times10^{20}$\,cm$^{-2}$. Thus, in the terminology of the \textsc{xspec} package, we fit the models \textsc{tbabs}$\,\times$\,\textsc{ztbabs}$\,\times$\,\textsc{zpowerlw} and \textsc{tbabs}$\,\times$\,\textsc{ztbabs}\,$\times$\,\textsc{zbbody}. The power-law provides an adequate fit to the data, with a w-stat of 197 for 245 degrees of freedom. The blackbody fit is a poor visual match and provides a far inferior fit with a w-stat of 359. 

The best-fitting photon index for our preferred power-law model is $\Gamma=1.54^{+0.14}_{-0.11}$. This is consistent with the lower end of the distribution for AGN, which have a population averaged $\langle\Gamma\rangle\approx 1.7-1.8$ \citep{Tozzi2006,Winter2009}. The fit does not require any intrinsic hydrogen column absorption. The model gives a mean unabsorbed flux $F_{\rm X}=1.21\times10^{-12}\,{\rm erg}\,{\rm cm}^{-2}\,{\rm s}^{-1}$. We also produced a light curve binned by visit, finding some indication of fading by a factor $\sim2$ between 2017 and 2018, however the significance is only $\approx 1\sigma$, and this is also comparable to the visit-to-visit scatter in the light curve. The count-rate light curve is given in Table \ref{tab:xray}, and the spectrum and light curve plots produced by the XRT online tools are provided in the appendix.

\subsection{Radio data}\label{sec:radio}

AT2017bcc was observed by the Jansky Very Large Array (VLA) over 4 logarithmically-spaced epochs between 2017-03-21 and 2018-01-17 from $3$-$25\,{\rm GHz}$ (PI: Alexander). These were reduced using the standard NRAO pipeline in Common Astronomy Software Applications \citep[CASA;][]{McMullin2007}. The observed fluxes are listed in Table \ref{tab:radio}. There is no obvious variability at any frequency, with typical flux of $\sim300\,{\rm \mu Jy}$. Historical observations from the VLA's FIRST and VLASS surveys contain detections with consistent flux density as early as 1993, suggesting that these emissions are unrelated to the recent optical flaring. If the radio emission were due to obscured on-going star formation, the implied star-formation rate (SFR) would be $\sim 9\,{\rm M}_\odot\,{\rm yr}^{-1}$, using the average flux at 5\,GHz with the relation from \citet{Yun2002}. Alternatively, this may indicate historical AGN activity on the timescale of thousands of years \citep{Hardcastle2020}.


\subsection{Spectra}

We obtained 20 epochs of spectroscopy between 2017-02-19 and 2023-01-15. Five epochs were taken with EFOSC2 on the New Technology Telescope (NTT) as part of ePESSTO+ and reduced via the PESSTO pipeline \citep{Smartt2015}. We took six with the OSMOS spectograph on the $2.4\,{\rm m}\:$ telescope at the MDM Observatory \citep{Martini2011}, three with ALFOSC on the NOT and six at the MMT Observatory (five with Bluechannel and one with Binospec) all of which were debiased, flat fielded, and cosmic ray corrected using either \textsc{iraf} or standard \textsc{Python} libraries. Finally, we acquired a spectrum with FORS2 on the Very Large Telescope (VLT) which was reduced using ESO Reflex. Relative flux calibration was achieved for all spectra using standard stars observed with the same instrument setups; calibrated spectra were then re-scaled to match contemporaneous photometry. Figure \ref{fig:all_spec} shows these spectra, further corrected for telluric absorptions and continuum subtracted to emphasise the emission lines.

\section{Host galaxy}\label{sec:host}

\begin{figure}
    \centering
    \includegraphics[width=\linewidth, trim=0cm 1cm 0.5cm 0.5cm]{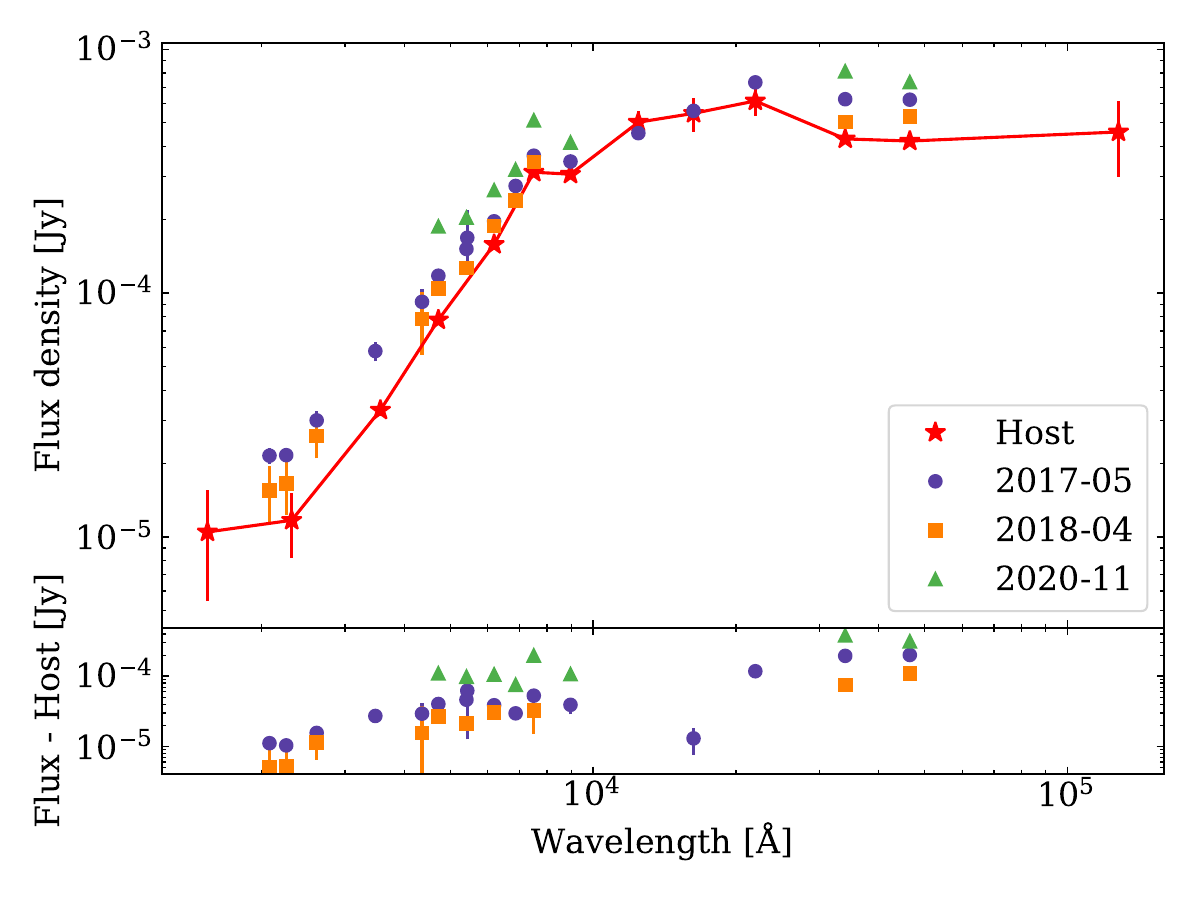}

    \caption{Spectral energy distribution of AT2017bcc at multiple epochs. Each point is an average flux from a 3-month time bin centred on the labelled epoch, plotted at the filter's effective wavelength. The archival SDSS, 2MASS, and WISE magnitudes are shown in red (labelled "Host"). The lower panel shows the same SEDs with the interpolated host magnitudes subtracted.}
    \label{fig:sed}
\end{figure}

We retrieved archival magnitudes of the host galaxy, SDSS J113152.98+295944.8 (or WISEA J113152.97+295944.9), from a number of catalogs. Optical data were obtained from the Sloan Digital Sky Survey \citep{Alam2015} in the $u,g,r,i,z$ bands. Near-infrared photometry was available from the 2 Micron All Sky Survey \citep[2MASS;][]{Skrutskie2006} in $J,H,K$, and mid-IR from the Wide-field Infrared Survey Explorer \citep[\textit{WISE};][]{Wright2010} in the bands $W1-3$. UV data was collected from the Galaxy Evolution Explorer \citep[GALEX;][]{Martin2005}. In all cases, we used the default reported magnitudes from each instrument.

We modelled the resulting spectral energy distribution using \textsc{Prospector} \citep{Johnson2021}. We used the Prospector-$\alpha$ model setup from \citet{Leja2017}, where the free parameters are the total mass formed in the galaxy, the metallicity, the current specific SFR in the last 50\,Myr, the widths in time of five equal-mass bins to model the star-formation history, three parameters that control dust absorption by both the interstellar medium and stellar birth clouds, and three parameters governing dust emission. We refer the reader to \citet{Leja2017} for an in-depth discussion of these parameters and their degeneracies. In particular, the default model does not work well for galaxies with a substantial AGN contribution, so we fit additionally for the fraction of IR emission originating from an AGN torus. The model posteriors were explored using \textsc{emcee} \citep{Foreman2013}. We ran the sampler with various combinations of 32-128 walkers \citep{Goodman2010} and chain lengths of $\sim1000-10,000$ steps. We found consistent posterior distributions, suggesting reliable convergence. The resulting corner plot is shown in Appendix \ref{sec:appendix}.

The resulting spectral energy distribution (SED) fit is shown in Figure \ref{fig:at2017bcc_sed_sdss}. This is a red galaxy with an old stellar population, with most star-formation occurring $\sim 9$\,Gyr ago, and a present-day SFR of $\sim 1\,{\rm M}_\odot\,{\rm yr}^{-1}$. The \textit{WISE} colours, $W1-W2=0.62\pm0.05$ and $W2-W3=1.92\pm0.34$, are most consistent with star-forming galaxies. This is in contrast to the majority of TDE host galaxies, which typically have star-formation histories peaking $\sim0.1-1$\,Gyr ago \citep{French2020}. The \textsc{Prospector} SFR is significantly lower than that implied by the radio luminosity of this galaxy. To investigate this issue further, we measured the average flux of the narrow [O II]\,$\lambda3727$ line in our spectra and used the calibration of \citet{Kennicutt1998} as an additional SFR indicator. This line was chosen as it is not blended with any of the complicated broad line profiles from the Balmer series. A luminosity of $\approx 6\times10^{39}$\,erg\,s$^{-1}$ implies a SFR of $\approx 0.1$\,M$_\odot$\,yr$^{-1}$. This is somewhat lower than the \textsc{Prospector} SFR, and much lower than the radio-implied SFR. Some of the discrepancy may be due to differences in aperture size between imaging and spectra, though we have attempted to scale our spectra to match in principle the apertures used for photometry. Our estimate from \textsc{Prospector} may therefore be an overestimate, and the true SFR likely lies somewhere between $\sim 0.1-1$\,M$_\odot$\,yr$^{-1}$. Given the very large discrepancy if we interpret it as star formation, we suggest that the radio emission is more likely related to historical AGN activity \citep{Hardcastle2020}.

The \textsc{Prospector} fit prefers a solar or slightly super-solar metallicity, $\log(Z/{\rm Z}_\odot)=0.1^{+0.07}_{-0.15}$, and a stellar mass $\log(M_*/{\rm M}_\odot)=10.70\pm0.08$. Assuming the mean bulge-to-total light ratio for this galaxy mass $B/T\sim0.7$ \citep{Stone2018}, this implies a SMBH mass of $\approx 1.5\times10^8\,{\rm M}_\odot$ using the relation of \citet{Kormendy2013}. An SMBH of this mass would disfavour a TDE origin for the variability in AT2017bcc, as this is above the Hills mass for direct capture (no disruption outside the event horizon) of a solar mass star \citep{Hills1975}, unless the BH is rapidly rotating and the star entered on a prograde orbit \citep{Leloudas2016}. However, the bulge mass of this galaxy is consistent with the most massive TDE host in \citet{Ramsden2022}, and their flatter BH-bulge relation for TDE host galaxies would suggest a SMBH more like $\sim10^7\,{\rm M}_\odot$.

Our fit does not require the existence of a powerful AGN prior to the onset of variability in 2017. The IR luminosity fraction is $f_{\rm AGN} = 0.08\pm0.02$, i.e. a contribution of $\lesssim10\%$ to the total IR emission could arise from a pre-existing dusty torus. However, this contribution is above zero at $4\sigma$, indicating that this structure likely does exist, even if it is not illuminated by a large accretion rate onto the central SMBH at that time. Overall, our SED model points to a largely dormant AGN with a SMBH mass in the range of a few $\times 10^7-10^8\,{\rm M}_\odot$.

\section{Photometric analysis}

\begin{figure}
    \centering
    \includegraphics[width=\linewidth, trim=0cm 1cm 0.5cm 0.5cm]{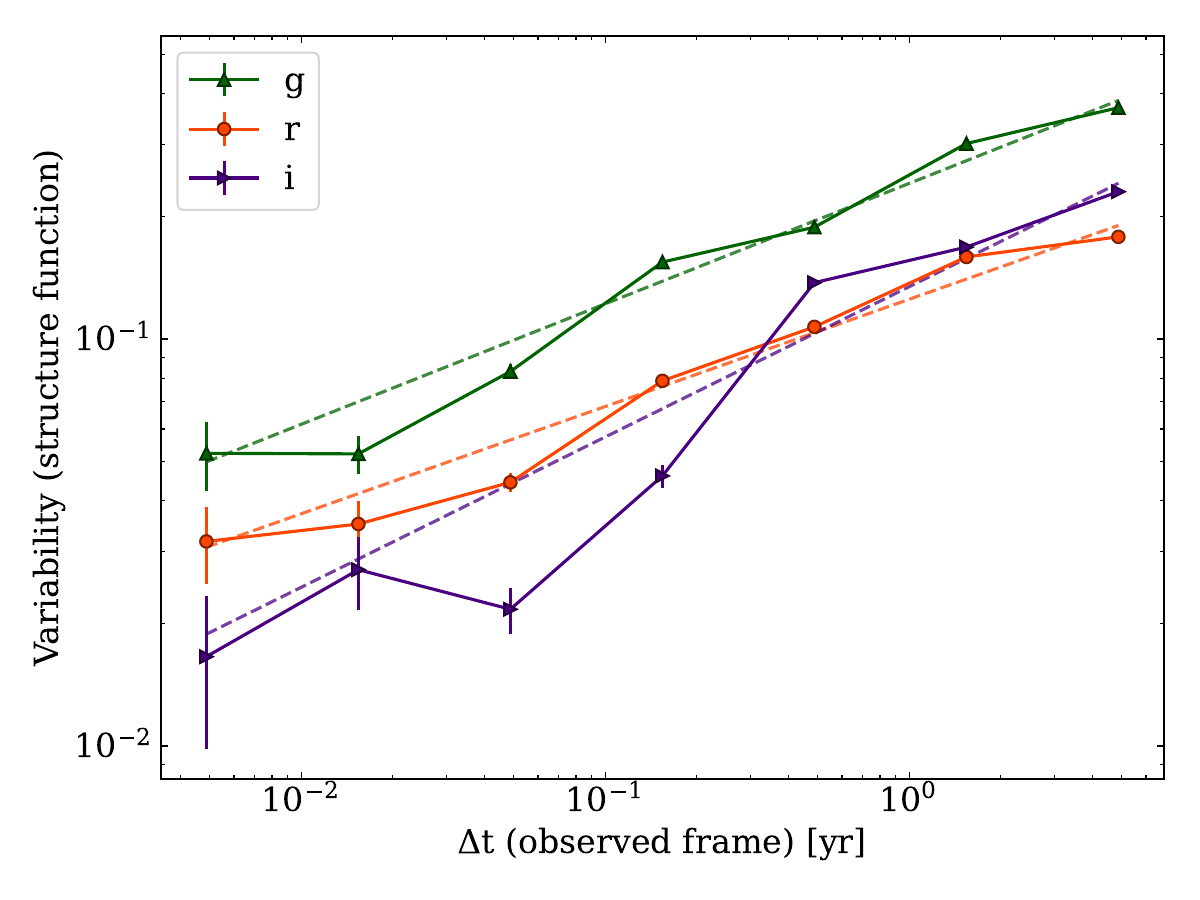}
    \caption{Structure functions, as defined in Equation \ref{eq:sf}, for AT2017bcc in $g$, $r$, $i$, using photometry from $2014$ onward with logarithmic time bins. The dashed lines show power-law fits to the structure functions, where $A$, the variability amplitude at 1 year, and $\gamma$, the power law exponent, are the parameters being optimised. The best-fit parameters for each band are as follows; $g: A=0.24, \gamma=0.30$; $r: A=0.13, \gamma=0.26$; $i: A=0.13, \gamma=0.37$.}
    \label{fig:sf}
\end{figure}

\begin{figure*}
    \centering
    \includegraphics[width=\linewidth, trim=0cm 1cm 0cm 0cm]{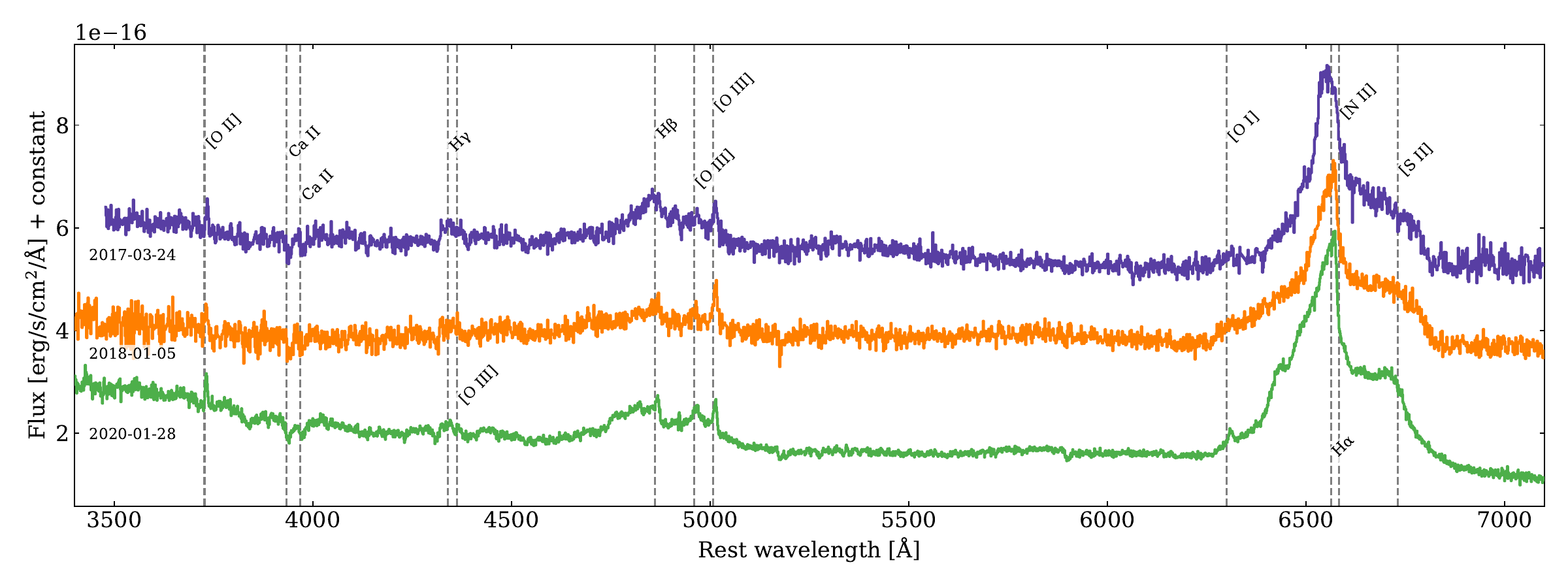}
    \caption{Three representative epochs of spectroscopy for AT2017bcc. All spectra were taken at the MMT observatory, the 2017 and 2018 epochs using Bluechannel, and the 2020 epoch using Binospec. All epochs have been flux-corrected using survey photometry and telluric-corrected. Relevant emission and absorption lines are marked with dashed vertical lines.}
    \label{fig:at2017bcc_specs}
\end{figure*}

\begin{figure*}
    \centering
    \begin{subfigure}{0.49\linewidth}
        \centering
        \includegraphics[width=\linewidth, trim=2cm 0cm 0cm 0cm]{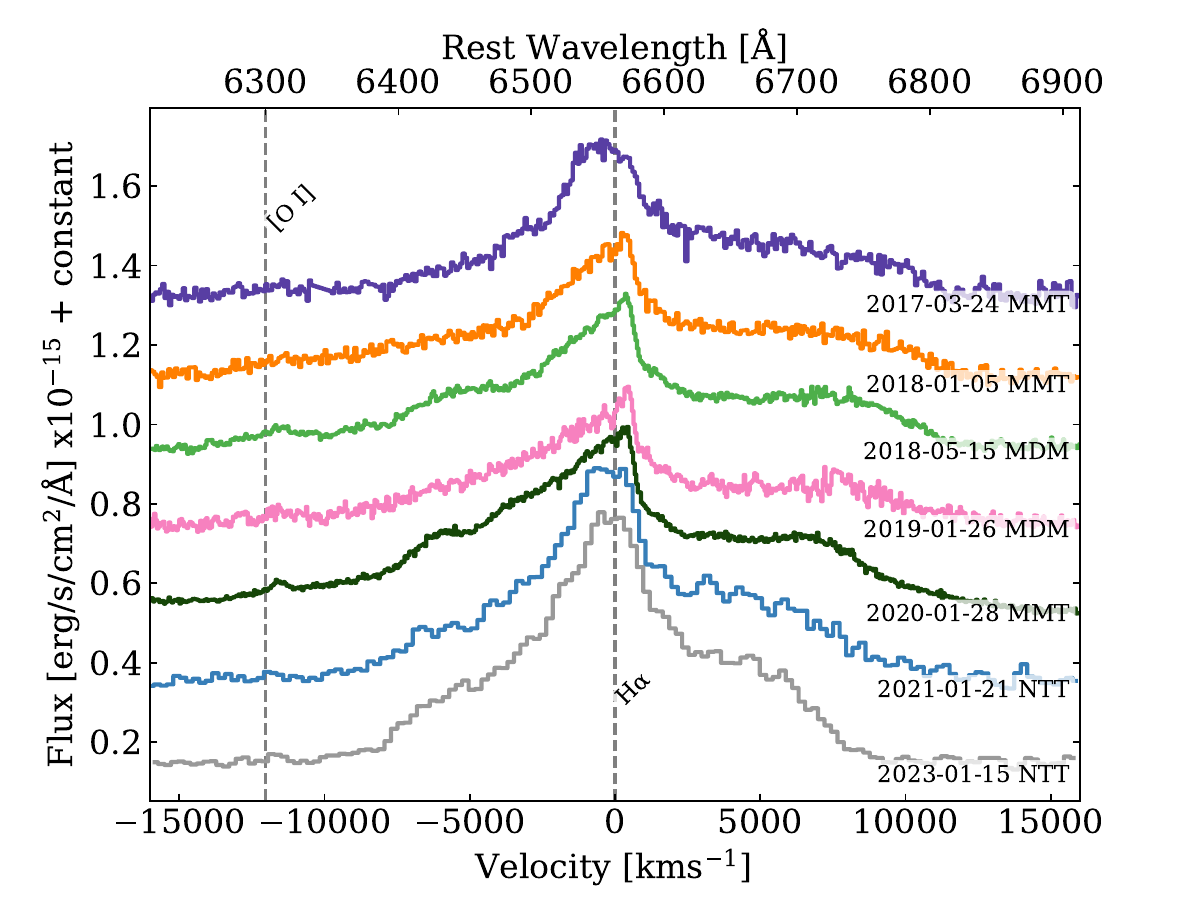} 
    \end{subfigure}
    \hspace{-1cm}
    \begin{subfigure}{0.49\linewidth}
        \centering
        \includegraphics[width=\linewidth, trim=0cm 0cm 2cm 0cm]{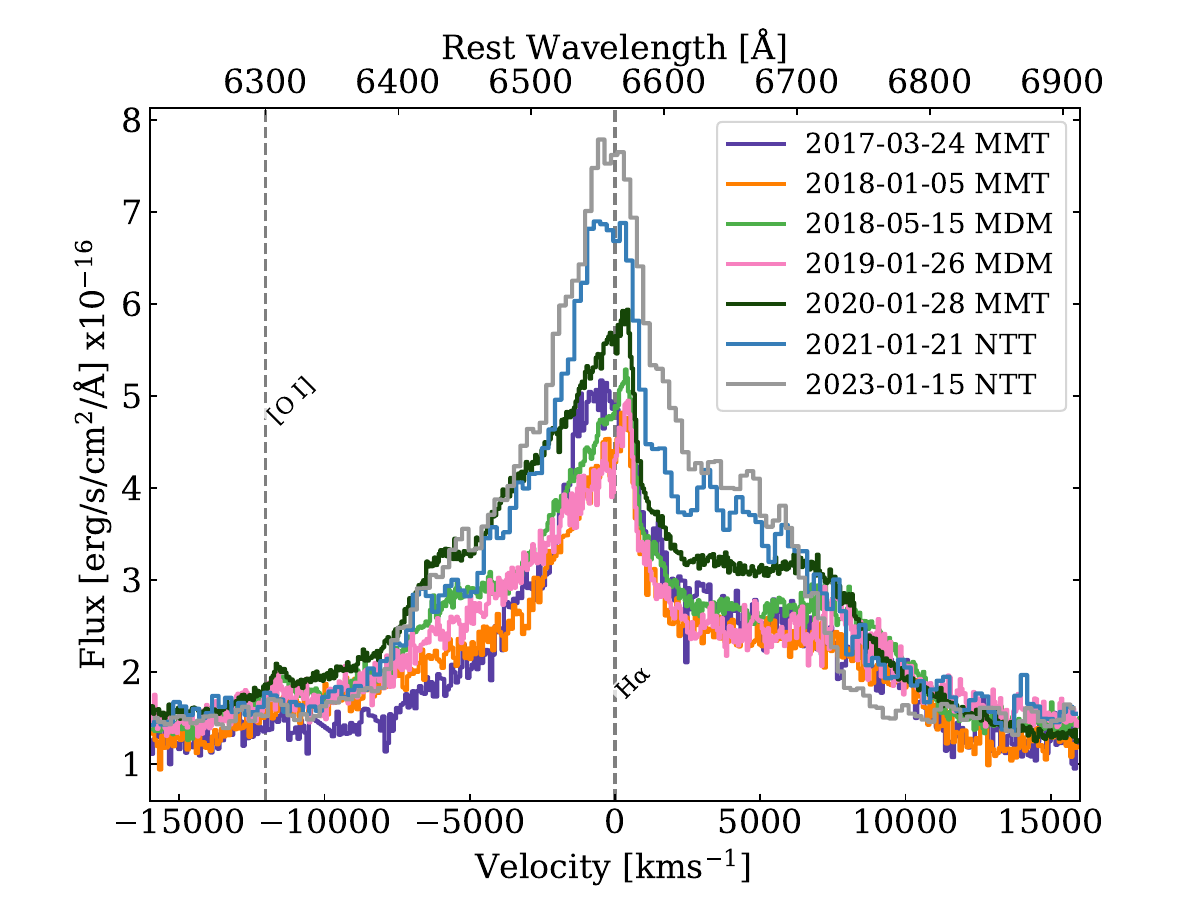} 
    \end{subfigure}

    \caption{Time-series evolution of the H$\alpha$ emission profile of AT2017bcc. All epochs have been flux-corrected using survey photometry and telluric-corrected. The left and right panels show the same spectra with and without vertical offsets respectively for visual clarity.}
    \label{fig:ha_stack}
\end{figure*}

\begin{figure*}
    \centering
    \begin{subfigure}{0.49\linewidth}
        \centering
        \includegraphics[width=\linewidth, trim=0cm 0.5cm 0cm 0cm]{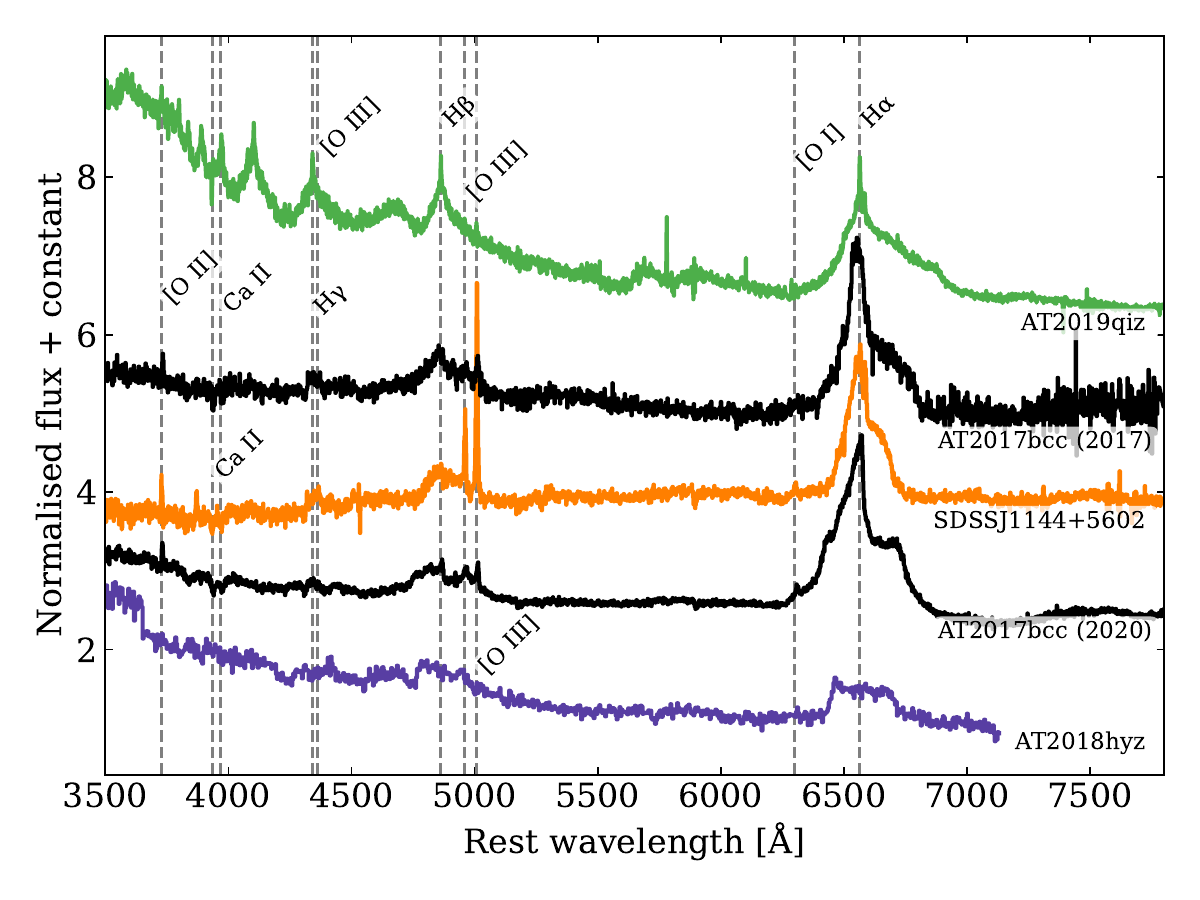} 
    \end{subfigure}
    \hspace{-0.2cm}
    \begin{subfigure}{0.49\linewidth}
        \centering
        \includegraphics[width=\linewidth, trim=0cm 0.5cm 1cm 1.8cm]{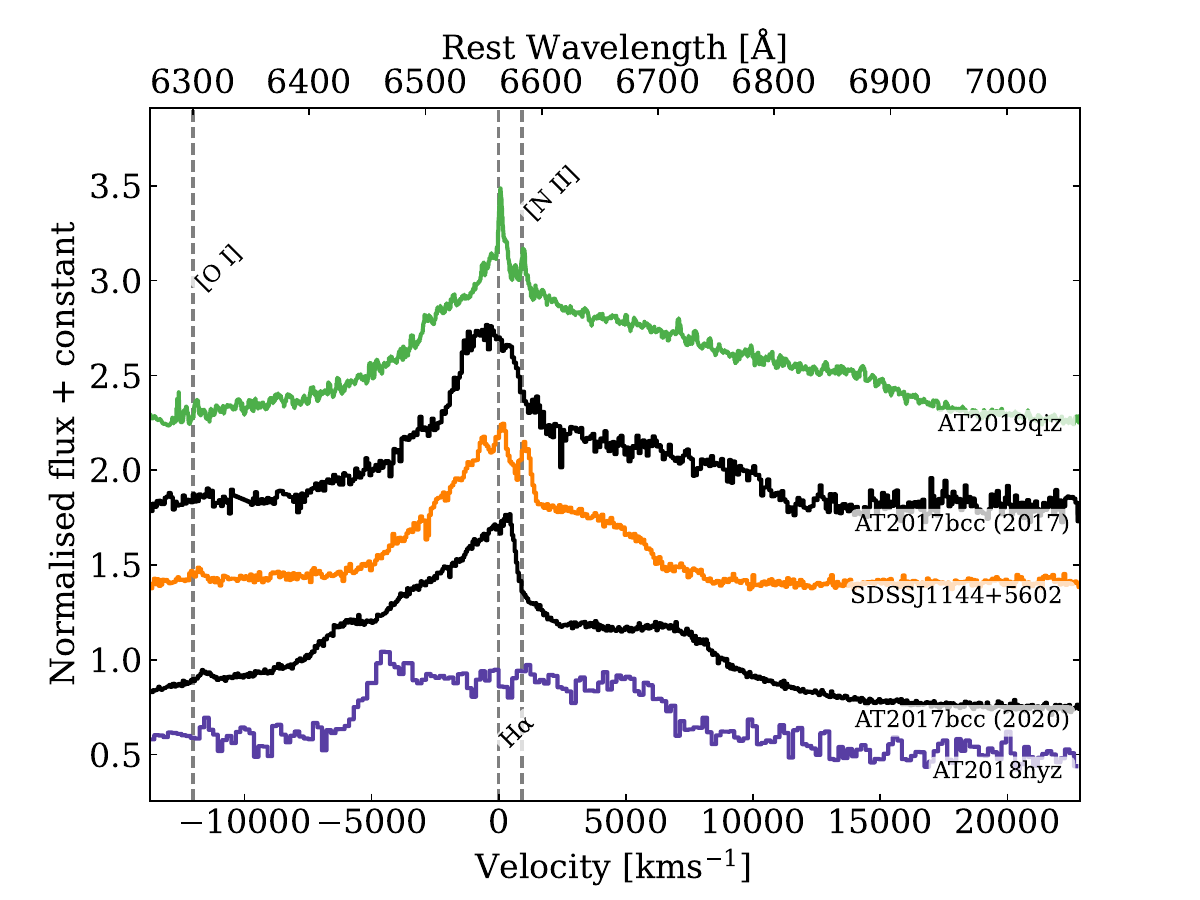} 
    \end{subfigure}

    \caption{Spectral comparison between two epochs of AT2017bcc (coloured black) and other nuclear transients. The comparison objects were selected from the literature for their similar features to AT2017bcc, and their high signal to noise ratios. All spectra have been normalised for visual clarity. The right panel shows a zoom-in on the H$\alpha$ profiles with alternative normalisation, for ease of comparison.}
    \label{fig:spec_comp}
\end{figure*}

\subsection{Light curve}

We have compiled a light curve for AT2017bcc consisting of over $15$ years of survey photometry with almost unbroken coverage, shown in Figure \ref{fig:at2017bcc_lc}. This photometry includes flux contributions from both the transient/variable nuclear source, and the host galaxy. We show in Figure \ref{fig:at2017bcc_hslc} a light curve with host light (from our \textsc{prospector} fit) subtracted. If we trust that the pre-flare photometry is free from contamination by any previous variability, the implied optical colour of the flare is apparently constant with $g-r\simeq0$. However, for the bulk of this analysis we will take the more agnostic approach of using the large aperture photometry without subtracting pre-flare light.

The light curve in Figure \ref{fig:at2017bcc_lc} shows two distinct phases separated by the discovery in early 2017. The first phase is one of quiescence, covered by SDSS, CRTS, and Pan-STARRS, which shows negligible change in magnitude over $\sim 10$ years. CRTS is particularly well sampled, showing little deviation ($\sigma = 0.1$ mag) from a $C$-band apparent magnitude of $18.2$ (absolute magnitude of $-21.0$). Two epochs of \textit{WISE} data in 2010 are also consistent with the first \textit{NEOWISE} magnitudes in 2014, both reporting $W1$ and $W2$-band magnitudes of $\sim17.4$ (AB), suggesting very little activity at both optical and infrared wavelengths. However, with sparser time sampling compared to the optical, it is harded to exclude variability pre-2017 in the IR.

The flare in early 2017 marks the beginning of a second, more active phase. This phase is well covered by ATLAS, \textit{NEOWISE}, and ZTF, which capture repeated flaring at irregular times and magnitudes. It is difficult to quantify the rise time, but the Pan-STARRS magnitudes in the $r$- and $i$-bands remain constant from the end of the \textit{CRTS} coverage until the first epoch of \textit{ATLAS} photometry. This constrains the optical rise to the $\sim200$ day gap before the peak in 2017. The first flare reaches a peak $o$-band magnitude of $17.5$ in ATLAS, marking a rise of $0.5$ mag from the previous year, and then fades back to a magnitude of $18.0$ in mid-2018. The object then spends the next two years slowly rising to a second peak in late 2020. Notably, the variability is more pronounced in bluer bands: while the $o$-band returns to its previous peak magnitude, the second $r$- and $g$-band peaks appear to be brighter than the first. This is then followed by a faster fade and re-brightening throughout 2021 and 2022, and then a third peak in 2023 exceeding the luminosity of the first two.

The second phase rules out a supernova origin for AT2017bcc, as supernovae are not known to re-brighten so dramatically on these timescales. Some TDEs are known to re-brighten after the initial peak, though they do so on timescales shorter than a year, and have not been observed to rise back to the original peak luminosity. Additionally, a third peak in the light curve is unprecedented for the known sample of TDEs \citep{Yao2023}. However, AGN do exhibit repeated flaring. Therefore, we investigate whether the statistical properties of the variability in AT2017bcc is consistent with an AGN.

\begin{figure}
    \centering
    \includegraphics[width=\linewidth]{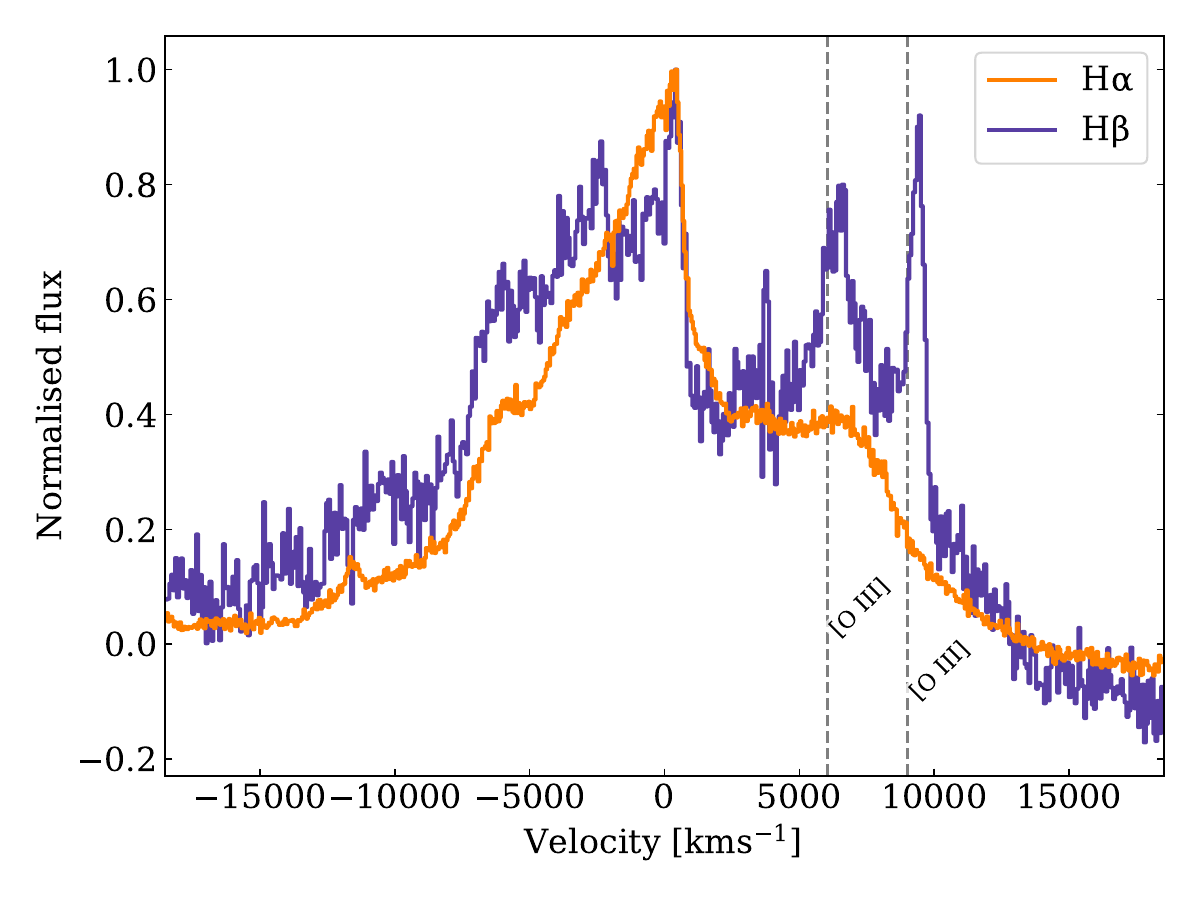}
    \caption{Overplot in velocity space of the H$\alpha$ and H$\beta$ regions of the 2020 epoch of spectroscopy. The two regions have been normalised to the same height for visual comparison. The narrow [\ion{O}{III}] emission lines are marked to distinguish them from the H$\beta$ profile.}
    \label{fig:ha_hb_overplot}
\end{figure}

\subsection{Structure function}

To compare the active phase of AT2017bcc's light curve to a typical AGN, we characterise its variability using the structure function, following the definition provided by \cite{Schmidt2010}. The variability $V$ in a given time bin $\Delta t$ is defined as
\begin{equation}\label{eq:sf}
    V(\Delta t) = \left\langle \sqrt{\frac{\pi}{2}} |\Delta m_{i, j}| - \sqrt{\sigma_i^2 + \sigma_j^2} \right\rangle_{\Delta t}\;,
\end{equation}
where $\Delta m_{i, j}$ is the difference in magnitude between observations $i$ and $j$, $\sigma_i$ and $\sigma_j$ are the photometric errors on those magnitudes, and $\langle \: \rangle_{\Delta t}$ signifies the average taken over all epoch pairs $i$, $j$ that fall in the bin $\Delta t$. Figure \ref{fig:sf} shows this function plotted for the $g$, $r$, $i$ bands using photometry from 2014 onward. This covers the epoch of variability, as well as the ambiguous period after the CRTS coverage.

We fit these structure functions with a power-law in the form $V = A \Delta t^\gamma$, where $A$, the variability amplitude at 1 year, and $\gamma$, the power law exponent, are the parameters being optimised. Focusing on the $r$-band to be consistent with \cite{Schmidt2010}, we find that $A=0.13, \gamma=0.26$. This lies within the quasi-stellar object (QSO) region of the parameter space described in \cite{Schmidt2010}. The caveat to this analysis is that we have very few cycles of the variability on the longer timescales, so these results may change over a longer observational baseline.

\subsection{Spectral energy distribution and IR evolution}

Due to the wide range of wavelengths probed by the photometry for this object, we are able to examine its UV-IR spectral energy distribution (SED) at multiple epochs. Figure \ref{fig:sed} shows the SED for AT2017bcc at the epochs of the first peak in luminosity, the first minimum, and just after the second peak. The source can be seen clearly above the host at all three epochs, with the largest flux excess in the UV and in the mid-infrared. These features are characteristic of outbursts from galactic nuclei, a population dominated by TDEs and AGN flares, with the mid-IR emission thought to originate from light echoes from pre-existing dust \citep{Mattila2018, Kool2020, Jiang2021, Reynolds2022}.

The IR luminosity begins to increase in $2015$/$2016$, with an eventual rise time on the scale of a year. The first peak in \textit{NEOWISE} occurs when the optical is already fading, potentially lagging the optical peak by $\sim6$ months. This implies a distance of $\sim10^{17}\,{\rm cm}$ between the optical and IR emission regions, consistent with the radius of an AGN dusty torus \citep{Hickox2018}. Taking the best-fit $W1$ and $W2$ flux from our \textsc{Prospector} model (consistent with the 2010 magnitudes) and subtracting these from the \textit{WISE} light curves, we see a possible colour change during the flare, becoming bluer in 2017, redder again during 2018, and evolving back towards $W1-W2=0$ during the second optical rise in 2019-2020. This could imply heating of dust by each flare. However, we caveat that the colour evolution is quite sensitive to the host subtraction, and the implied blackbody temperature of the dust emission is at all times in the range $\approx 900\pm150$\,K.

We can estimate the bolometric luminosity needed to heat the dust following \citet{Somalwar2022}. For a dust temperature of 900\,K and a radius of $10^{17}$\,cm and an assumed grain size of $0.1\mu$m \citep{Draine1984}, their equation 4 leads to $L_{\rm bol}\sim {\rm few}\times 10^{46}$\,erg\,s$^{-1}$ assuming a covering factor of 1, or $\sim 10^{45}$\,erg\,s$^{-1}$ for a more realistic covering factor of $\sim0.1$ \citep{Ricci2017}.

\section{Spectroscopic analysis}
\subsection{Spectral evolution and comparisons}

\begin{figure*}
\begin{subfigure}[t]{0.35\textwidth}
    \includegraphics[width=\textwidth]{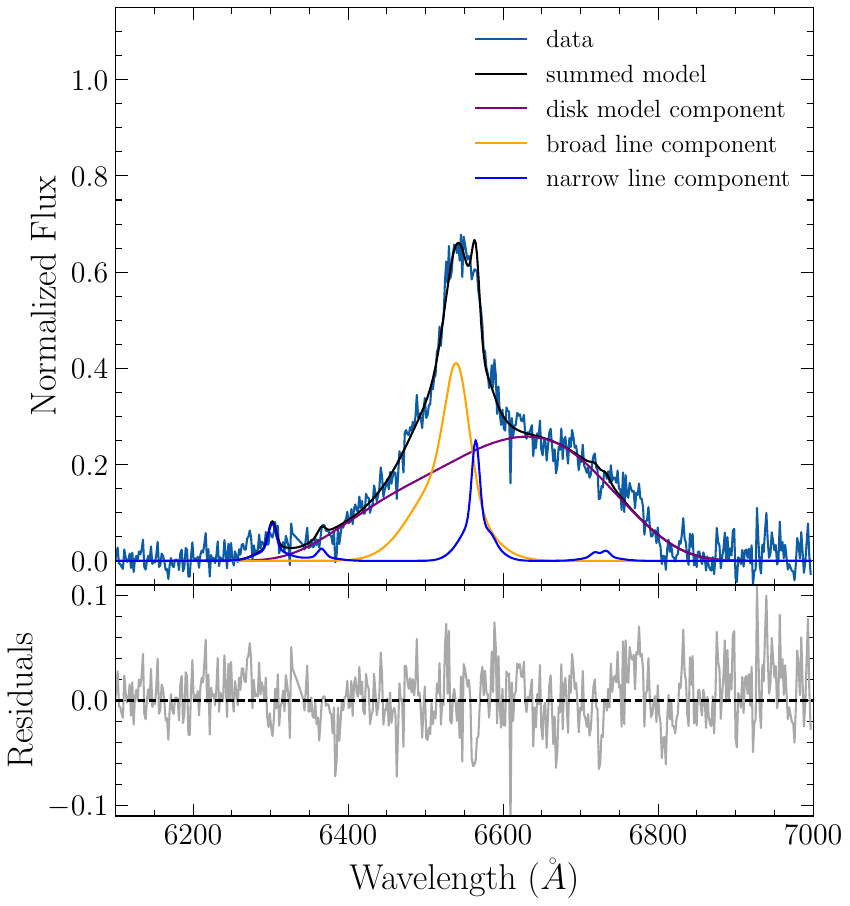}
    \caption{2017-03-24}
\end{subfigure}
\begin{subfigure}[t]{0.35\textwidth}
    \includegraphics[width=\linewidth]{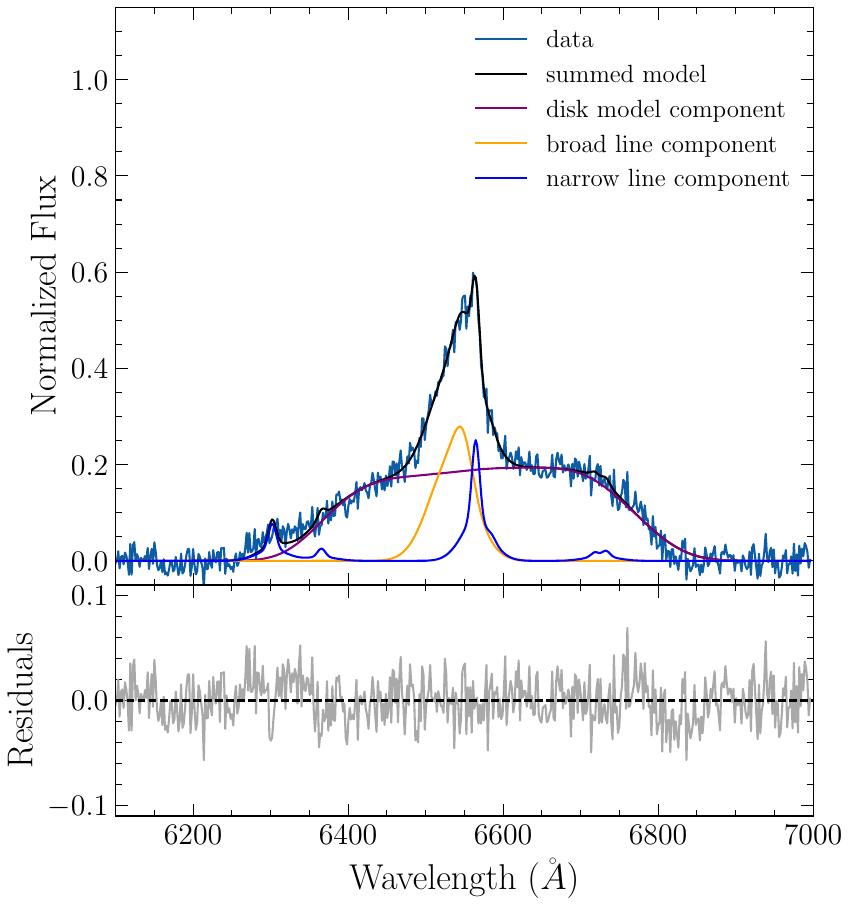}
    \caption{2018-01-05}
\end{subfigure}
\begin{subfigure}[t]{0.35\textwidth}
    \includegraphics[width=\linewidth]{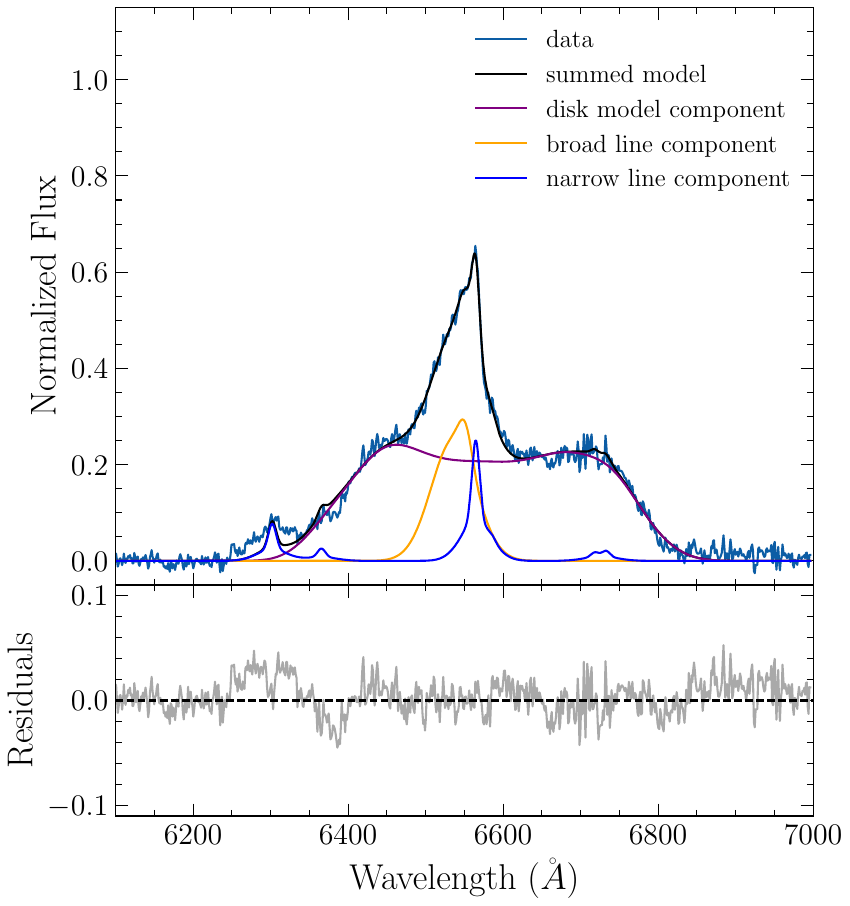}
    \caption{2018-05-15}
\end{subfigure}
\begin{subfigure}[t]{0.35\textwidth}
    \includegraphics[width=\linewidth]{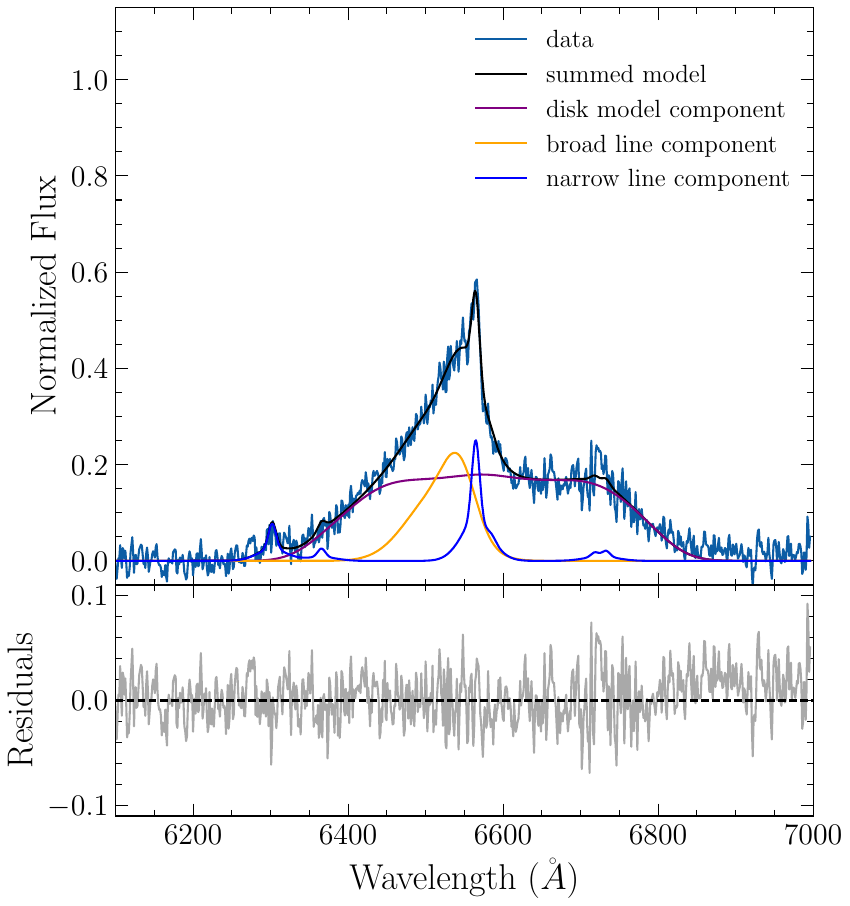}
    \caption{2019-01-26}
\end{subfigure}
\begin{subfigure}[t]{0.35\textwidth}
    \includegraphics[width=\linewidth]{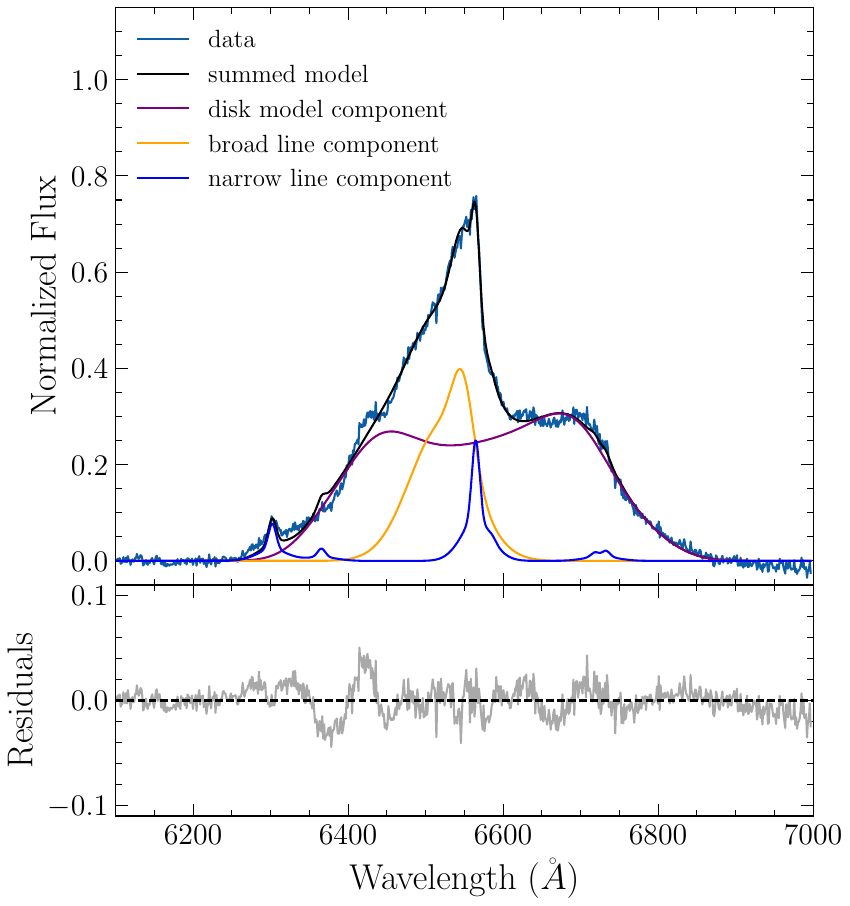}
    \caption{2020-01-28}
\end{subfigure}
\begin{subfigure}[t]{0.35\textwidth}
    \includegraphics[width=\linewidth]{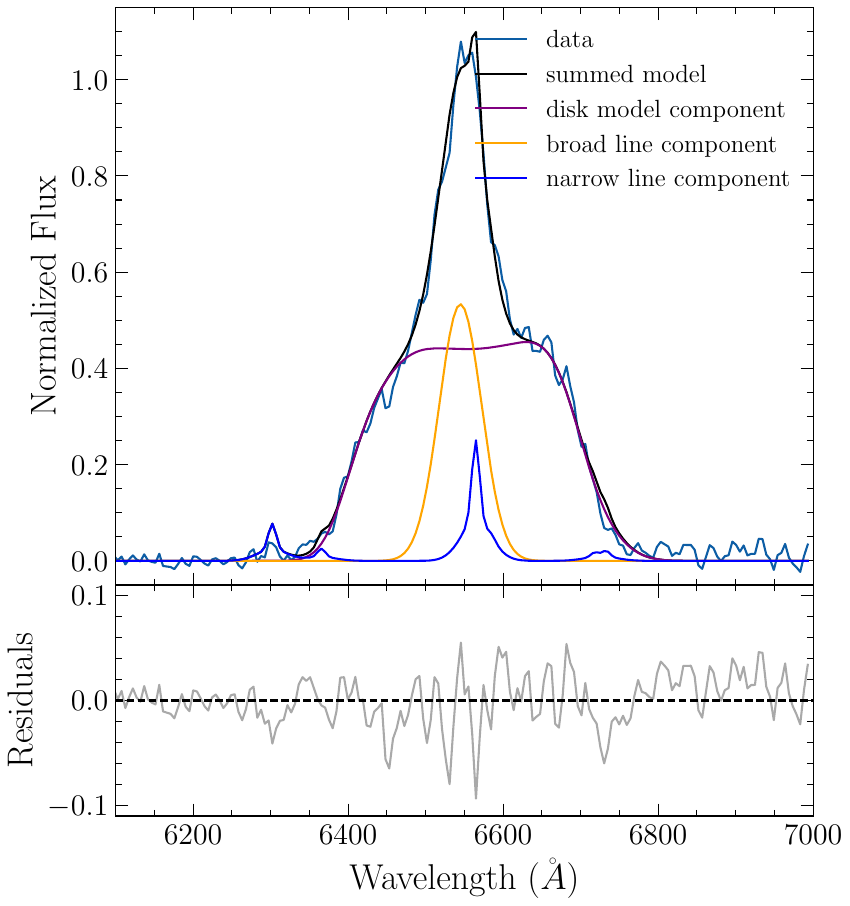}
    \caption{2023-01-15}
\end{subfigure}
\caption{Examples of the best-fit H$\alpha$ disc and broad line models for 6 different spectral epochs of AT2017bcc. The fits were made to continuum-subtracted and telluric-corrected spectra. We have plotted the $1\sigma$ uncertainty bands for the best-fit model given the distributions of the parameters found by the MCMC modeling, but we note that the range of feasible fits is very narrow for our selected model. The residuals after subtracting both models and the narrow-line component are shown below each profile. A diagram of the emission regions is shown in Appendix \ref{sec:appendix}.}
\label{fig:specfits}
\end{figure*} 

\begin{figure*}
\begin{subfigure}[t]{0.45\textwidth}
    \includegraphics[width=\textwidth]{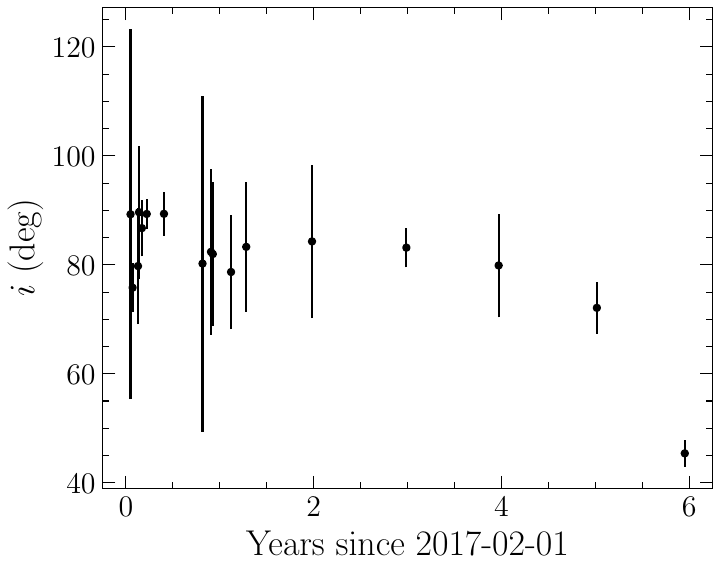}
    \caption{Disc inclination angle}
\end{subfigure}
\begin{subfigure}[t]{0.45\textwidth}
    \includegraphics[width=\linewidth]{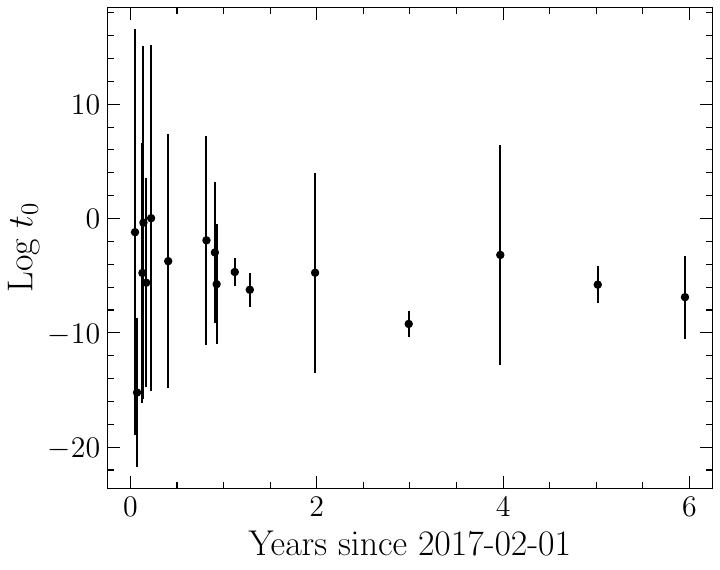}
    \caption{Wind amplitude}
\end{subfigure}
\begin{subfigure}[t]{0.45\textwidth}
    \includegraphics[width=\linewidth]{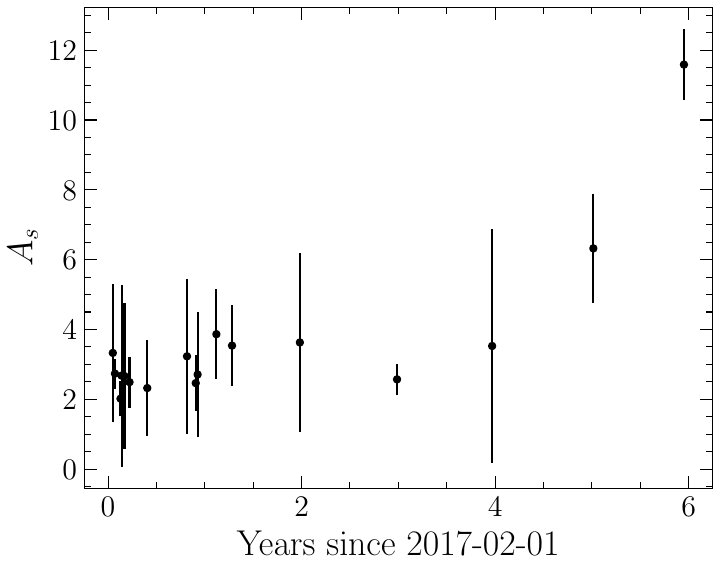}
    \caption{Spiral arm amplitude}
\end{subfigure}
\begin{subfigure}[t]{0.45\textwidth}
    \includegraphics[width=\linewidth]{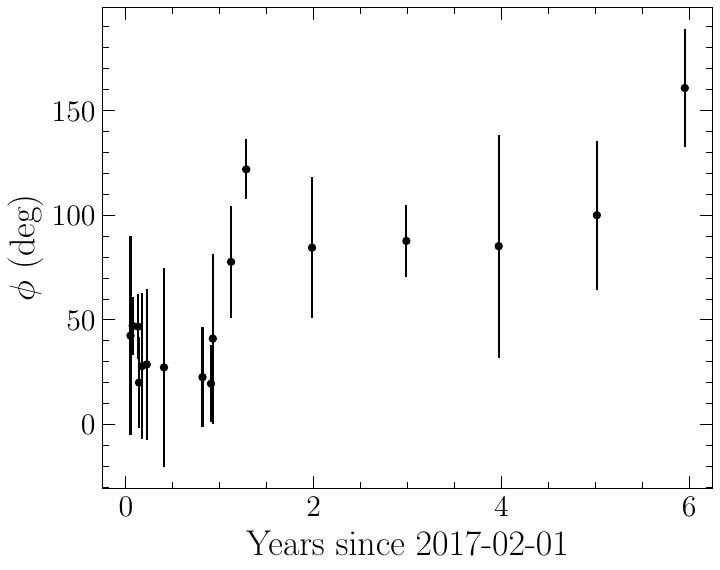}
    \caption{Spiral arm phase}
\end{subfigure}
\caption{Time series of 4 key parameters from fitting the optical emission profile of AT2017bcc, described in Section \ref{sec:profile-fitting}. These show the inferred evolution of the accretion disc's inclination angle, wind, and spiral arm.}
\label{fig:timeseries}
\end{figure*}

\begin{figure}
    \centering
    \includegraphics[width=\linewidth]{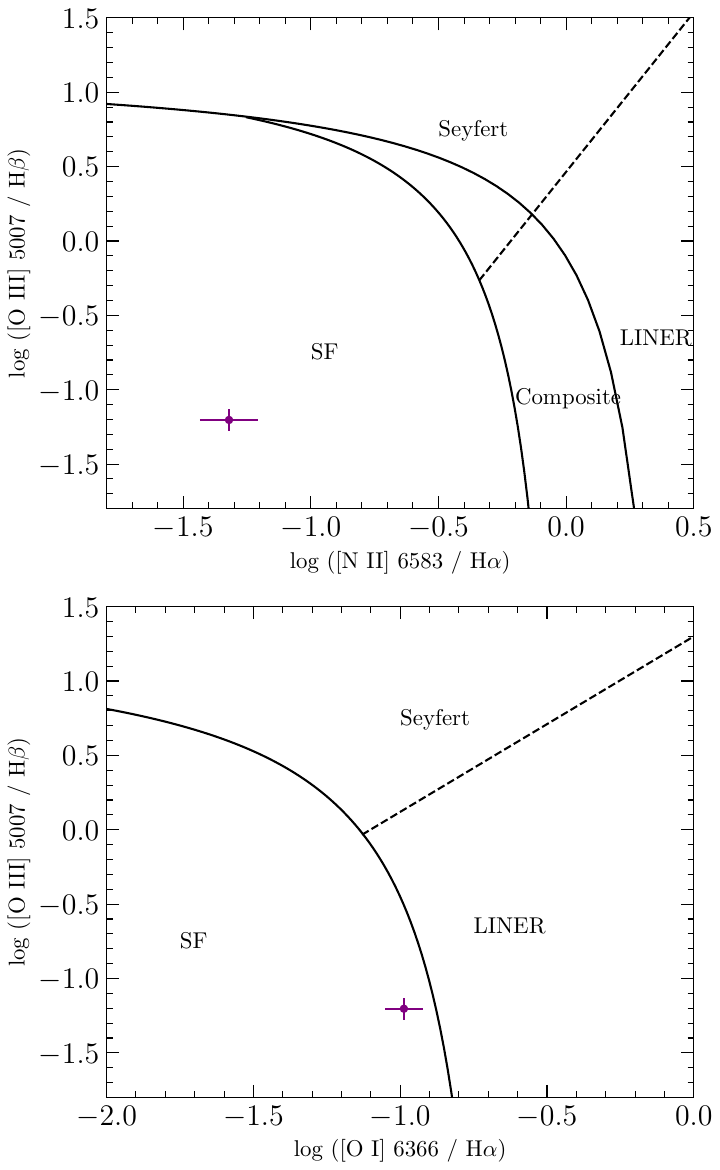}
    \caption{BPT diagrams showing the best-fit narrow emission line ratios fit from modelling of the high S/N 2020-01-28 spectrum.}
    \label{fig:bpt}
\end{figure}

The optical spectra of AT2017bcc show a blue continuum and broad, multi-component hydrogen emission. Figure \ref{fig:at2017bcc_specs} presents three representative epochs of spectroscopy: shortly after discovery, during the first minimum, and during the rise to the second peak in luminosity. 

The most notable difference between these epochs is the evolution of the H$\alpha$ line profile. The profile in 2017 shows two main components: an asymmetric, very broad component with an excess in flux on the red shoulder, and a narrower component which appears to be blue-shifted by $\sim 500 \, {\rm kms^{-1}}$. In 2018, after the initial flare has faded, the narrower component becomes distinctly asymmetric with a blue wing, and peaks at the rest wavelength of $6568\,$\AA. As the luminosity rises to a second peak in 2020, the broad component develops a blue shoulder as the profile becomes overall `boxy' in shape.

Figure \ref{fig:ha_stack} shows a zoom-in on H$\alpha$ and its subsequent development over the next $3$ years. Between 2020 and 2021, the red shoulder of the broad component becomes stronger, such that the two wings are once again unequal in height. The relative heights of the blue and red shoulders do not appear to evolve further past 2021. Unfortunately, the spectral resolution of the data after 2020 is not sufficient to determine whether the narrower component is still asymmetric or simply blueshifted.

Figure \ref{fig:spec_comp} compares the 2017 and 2020 epochs of AT2017bcc to optical spectra from other nuclear transients. The early-time broad component of H$\alpha$ resembles that of SDSS J1144+5602, a double-peaked AGN \citep{Paris2017}, though the shape of the wings is more similar to those of AT2019qiz, a nearby TDE \citep{Nicholl2020,Hung2021}. This same broad component in the later profile has much steeper sides, bearing more resemblance to the H$\alpha$ profile of AT2018hyz, a unique TDE with double-peaked emission lines \citep{Short2020, Gomez2020,Hung2020}. In the case of AT2018hyz, this broad emission profile has been attributed to an exposed accretion disc.

The narrow component in the later spectrum has a steep drop in flux on the red side, and a shallower slope on the blue side. This is most similar to the central component and blue wing of H$\alpha$ in SDSS J1144+5602 (though the asymmetry is much more pronounced in AT2017bcc). These comparisons may suggest that the H$\alpha$ profile of AT2017bcc is comprised of a broad, accretion disc-like component similar to some TDEs as well as double-peaked AGN \citep{Eracleous1994}, which has evolved over time; and a narrower, asymmetric "shark fin" component similar to some double-peaked AGN.

The broad, double-peaked component can also be seen in the H$\beta$ profile, with the same width in velocity space as in H$\alpha$, as shown in Figure \ref{fig:ha_hb_overplot}. There is also an asymmetry in the central component of the H$\beta$ profile, though its blue shoulder is less smooth. Unlike most AGN, the narrow [\ion{O}{III}] emission at $4959\,$\AA$\!$ and $5007\,$\AA$\!\,$ is weak. The [\ion{O}{III}] flux appears comparable to the narrow component of H$\beta$. As seen in SDSSJ1144+5602, AGN typically have an [\ion{O}{III}] / H$\beta$ flux ratio of $\gg 1$ \citep{Stern2012}.

\subsection{Line profile fitting} \label{sec:profile-fitting} 
We applied the circular accretion disc model from \cite{Chen1989} to the double-peaked spectra of AT2017bcc to determine the time-varying accretion disc properties. We fit to normalised spectra, and so this analysis is not sensitive to precise flux calibrations of spectra obtained with different slit widths. We first modelled the high S/N continuum-subtracted spectrum from 2020-01-28 with the following components: a circular accretion disc model for the H$\alpha$ and H$\beta$ broad emission line regions; the narrow emission lines from H$\alpha$, H$\beta$,  [S\,{\sc ii}] $\lambda$6717, 6731, [N\,{\sc ii}] $\lambda$6550, 6575, [O\,{\sc i}] $\lambda$6302, 6366 and [O\,{\sc iii}] $\lambda$5007, 4959; and a broad two Gaussian component close to the H$\alpha$ and H$\beta$ narrow lines to describe the shark fin feature. 

The disc models had $2$ parameters in common for both the H$\alpha$ and H$\beta$ emission regions: inclination angle $i$ where 0 degrees is face-on and 90 degrees is edge-on, and a local turbulent broadening parameter $\sigma$\ (km/s). Three disc parameters were allowed to differ for each of the H$\alpha$ and H$\beta$ emission regions: the emissivity power law index $q$, and the inner and outer dimensionless gravitational radii of the disc $\xi_1$ and $\xi_2$. The simple circular disc model did not adequately describe the flat red shoulder of the 2018-2023 spectra, so we enabled a wind component to the model to increase the `boxiness’ of the double-peaked profile \citep{Nguyen2018}. The disc wind had $3$ free parameters: the opening angle of the wind $\theta$, the wind optical depth $\tau$, and the optical depth normalisation $t_0$ which affects the strength of the wind \citep{Murray1996,Flohic2012}. Finally, we enabled a single spiral arm in the accretion disc with free parameters: amplitude $A_{\rm s}$ (expressed as a contrast ratio relative to the rest of the disc), orientation angle $\phi$ (deg), width $w$ (deg), and pitch angle $\psi$ (deg). This was required to describe the flux ratio of the red and blue shoulders being $>$1, as is common amongst disc emitters (e.g. \cite{Storchi2003}).

The narrow lines were fitted with respect to 5 narrow line flux ratio parameters: [N\,{\sc ii}] $\lambda$6583 /H$\alpha$,  [S\,{\sc ii}] $\lambda$6731/H$\alpha$, [O\,{\sc i}] $\lambda$6366/H$\alpha$, [O\,{\sc iii}] $\lambda$5007/H$\beta$; H$\alpha$/H$\beta$. The [N\,{\sc ii}], [S\,{\sc ii}], [O\,{\sc i}], and [O\,{\sc iii}] doublet flux ratios were fixed to theoretical values of 2.95, 1.3, 0.33, and 2.88 respectively. The narrow lines were described by two component Gaussians of the same central wavelength with 3 free parameters which were common for all narrow lines: the width of the first Gaussian component $\sigma_1$, the width of the second Gaussian component $\sigma_2$, and the flux ratio of the two components $f_1/f_2$. A two component Gaussian was chosen because a single Gaussian did not sufficiently describe the peaky line shapes of the [O\,{\sc iii}] narrow lines which are less contaminated by the disk profile. The redshift of the lines was fit as a single parameter so that the line centroids could be optimised while being tied to the same redshift. The shark fin feature was fitted with a double Gaussian, with each Gaussian having 2 free parameters which are separate for H$\alpha$ and H$\beta$: the width $\sigma_b$ and the rest wavelength offset from the narrow H$\alpha$ or H$\beta$ line central wavelength $\Delta$. The optimal amplitudes for the sets of narrow lines and broad lines were obtained by solving the covariance matrix for a given set of narrow line, broad line and disc model components, so they were not fitted as parameters during optimization. 

We first found a reasonable initial fit using the least-squares optimisation implemented in \textsc{Python} using the \textsc{scipy} package. We then explored the posteriors using \textsc{emcee} \citep{Foreman2013} with 60 walkers initialized at the best-fit values from the least-squares fit, distributed according to the 1$\sigma$ error found from the least-squares covariance matrix. The emcee fitting was run for 5000 iterations with a burn-in time of 4000 iterations.  

For the remainder of the spectra, we fixed the narrow line shapes and amplitudes to the best-fit values found from the 2020-01-28 spectrum, but left all broad-line and disc parameters free for re-fitting. The fitting procedure was repeated using the best-fit values from \textsc{emcee} fitting of the 2020-01-28 spectrum to initialize the walkers, but with only 2500 iterations and a burn-in time of 2000 iterations, which was the number of iterations required for the walkers to stabilise at their optimal values.  

All epochs were well-described by our disc$+$two-Gaussian central broad-line model. The fits to the H$\alpha$ regions of the spectra are shown in Figure \ref{fig:specfits} and the fits to the H$\beta$ regions of the spectra are shown in Figure \ref{fig:specfitsbeta}. The best-fit H$\alpha$ disc parameters are shown in Table \ref{tab:discparams} and the best-fit shark fin parameters are shown in Table \ref{tab:outflowparams}. We note that the absolute flux of the central shark fin structure and the disc profile increased over time relative to the narrow line fluxes. The full width half maximum (FWHM) of the central outflow structure was approximately 70\ \AA\ and showed some changes in morphology through the 6 years of data (Figure \ref{fig:specfits}). Evolution in the shape of the shark fin profile, as described by the free parameters for the two-Gaussian model in Table \ref{tab:outflowparams}, contributed substantially to changes in the appearance of the left shoulder of the disc profile. Most disc parameters did not evolve substantially over the 6 years of data, and changes to the width and boxiness of the profile from 2018--2023 could be accounted for solely by changes to the inclination angle and the spiral arm strength and phase. 

The disc profiles had a constant emissivity power law index of $q\sim1.75$ for both H$\alpha$ and H$\beta$ emission regions and a constant turbulent broadening parameter of $\sigma \sim1267$\ km/s. The inner radius stayed constant at $\xi_1 \sim 700$ for the H$\alpha$ emission region while the outer radius was constant at $\xi_2 \sim 2000$. The best-fit inclination angle of the disc stayed within the range of $75<i<90$\textdegree\ in 2017 before decreasing further to $i\sim45$\textdegree\ in later epochs, accounting for the decrease in the width of the double-peaked profile (Figure \ref{fig:timeseries} a). 

A modest wind of opening angle $\theta \sim0.8$\textdegree\, and optical depth $\tau \sim 0.6$ was found to improve the fit to the boxy disc profiles over all epochs (Figure \ref{fig:timeseries} b). The spiral arm parameters evolved over time, with the phase of the arm evolving from $\phi\sim25$ to $\phi\sim 160$ and the amplitude increasing from $A\sim2.5$ to $A\sim11.5$ in later epochs when it became more essential to describe the relative flux of the red and blue peaks of the double-peaked profile (Figure \ref{fig:timeseries} c). The gradual variation in these parameters over multi-year timescales matches the variability timescales reported in other double-peaked emitters \citep{Schimoia2017}. The spiral arm had a best-fit width of $\sim 50$\textdegree, and the pitch angle of the arm was approximately $\sim 25$\textdegree. 

The narrow line ratios were well within the star-forming regions of the BPT diagram indicating no presence of long-lived AGN activity (Figure \ref{fig:bpt}). We show only the [N\,{\sc ii}] $\lambda$6583 /H$\alpha$ and [O\,{\sc i}] $\lambda$6366/H$\alpha$ BPT diagrams because the [S\,{\sc ii}] $\lambda$6731/H$\alpha$ fitting may be subject to contamination from the edge of the red disc shoulder. The implied luminosity of the H$\alpha$ line in our fits is $\sim10^{40}$\,erg\,s$^{-1
}$, which would correspond to a SFR $\sim 0.1$\,M$_\odot$\,yr$^{-1}$, in agreement with the [O II] line measurement in section \ref{sec:host}, and somewhat lower than the SFR rate estimated from host SED modelling. This suggests that recent AGN activity does not have a large effect on the narrow line fluxes, since a contribution from AGN excitation could only serve to elevate the narrow line fluxes above those expected based on other SF indicators.

In order to see if a disc profile alone could be responsible for both the double-peaked structure and the central shark fin structure, we attempted to fit an alternative model for the 2020-01-28 spectrum which lacked the additional multi-Gaussian broad line for the shark fin component. This did not produce a high quality fit because the circular disc model was unable to account for both the shark fin feature and the blue shoulder of the double-peaked spectrum without the additional multi-Gaussian component.

We caution that the various parameters needed to describe the complex line profile may lead to inevitable degeneracies, especially given the cross-contamination of the evolving shark fin-shaped broad line and the double-peaked disc profile. We also note that it is possible that the two component Gaussian used to model the narrow lines may be compensating for imperfections in the disk model. However, this should have been mitigated by the requirement that the narrow line morphologies must be consistent across all narrow emission lines, including those which are less contaminated by the disk profile such as [O\,{\sc iii}]. We also note that the narrow line fluxes and morphologies were fixed after the initial fit to the 2020-01-28 spectrum, and in all cases the disc and outflow model was capable of accurately describing the evolving parts of the spectrum.

In summary, our modeling of the H$\alpha$ and H$\beta$ broad-line regions finds that the apparent changes to the spectrum over time can be accounted for by: changes to the relative flux of the broad line and narrow line components, changes to the shark fin morphology, and gradual changes to the disc inclination angle along with the spiral arm location and amplitude.

\section{Discussion}

\subsection{Presence of an active SMBH}
The repeated re-brightening of its light curve rules out a supernova or a `typical' TDE as the origin of AT2017bcc. After the original flare in 2017, the luminosity faded for a year, and then rose again to match the first peak, and even exceeded this later on. Even if the first flare was caused by a one-off event, like a TDE, the years-long rise to the later peaks would far exceed the fall-back time for a plausible disruption \citep{Guillochon2013,Coughlin2022}. Although repeating partial TDEs are possible and may have been observed \citep{Payne2021,Wevers2023}, the light curve of AT2017bcc appears stochastic rather than varying on an orbital period. The most plausible cause of this slower evolution is a gas reservoir in the form of an accretion disc around the central SMBH, i.e.~a pre-existing but dormant AGN.

This scenario is also supported by a statistical treatment of the light curve. We measured the variability of the recent optical emission from AT2017bcc using its structure function. Power-law fits to this structure function showed best-fit parameters for the $r$-band of $A=0.13, \gamma=0.26$. This is well within the region defined in \cite{Schmidt2010} which separates quasars from other variable sources.

We also find similarities between the spectra of AT2017bcc and those of double-peaked AGN, especially at later times. Once the initial flare had faded the H$\alpha$ emission profile developed a distinctly asymmetric peak resembling a "shark fin". This unusual profile is also seen in SDSS J1144+5602, a known quasar with double-peaked emission features \citep{Paris2017}.

Double-peaked AGN are a subclass of AGN, spectroscopically selected for their unique emission features. They make up a significant portion of known AGN, accounting for $19\%$ of those observed with ZTF \citep{Ward2023}. Half of this sample show dramatic evolution in the heights of the red/blue peaks in their H$\alpha$ profiles over time, attributed to the migration of hotspots in the accretion disc. This kind of evolution can also be seen in the wings of AT2017bcc's H$\alpha$ profile, further implying that it is currently behaving as a double-peaked AGN. These objects typically host higher mass SMBHs and accrete at larger Eddington ratios, and are more likely to be X-ray and radio bright \citep{Ward2023}. AT2017bcc seems to represent the discovery of a double-peaked AGN undergoing a dramatic change in its observed properties after a long period of quiescence.

\subsection{Narrow line emission}

\begin{figure*}
    \centering
    \includegraphics[width=\linewidth]{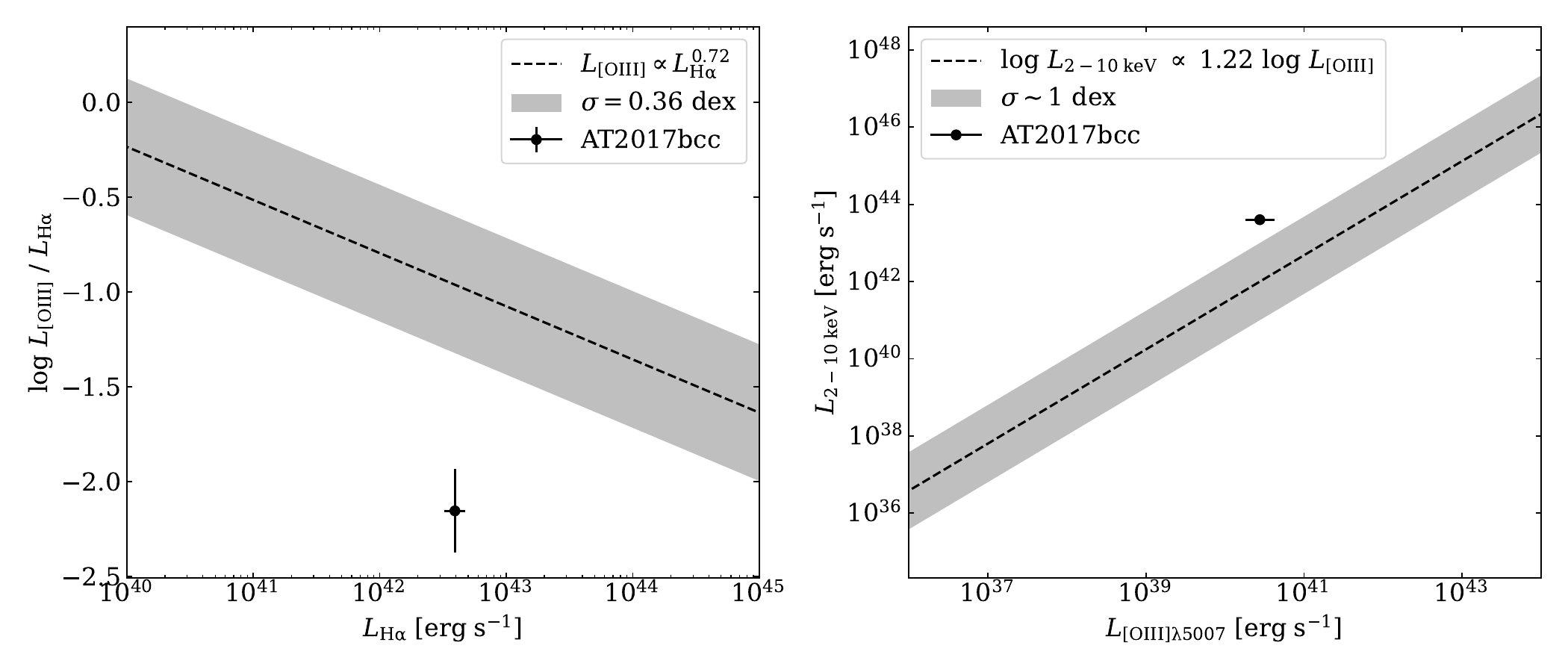}
    \caption{Distributions of [\ion{O}{III}] to H$\alpha$ (left) and X-ray (right) luminosities for AGN, with AT2017bcc marked in black. The distribution in H$\alpha$ luminosity is shown in \protect\cite{Stern2012}, where they observe a scatter of $\sigma = 0.36$ dex. The distribution in X-ray luminosity is described in \protect\cite{Panessa2006}, from which we infer a scatter of $\sigma \sim 1$ dex.}
    \label{fig:o3_emission}
\end{figure*}

Narrow-line [\ion{O}{III}] emission is a defining characteristic of AGN spectra. The strength of [\ion{O}{III}] emission around $\sim 5000\,$\AA$\!$, relative to emission at other wavelengths, is used to categorise active galaxies. One of the most common distinctions is that between high-ionisation Seyferts and low-ionisation nuclear emission line regions (LINERs) \citep{Stern2013}. These groups are usually separated by their position on BPT diagrams \citep{Baldwin1981}, which compare [\ion{O}{III}] line strength to [\ion{N}{II}], [\ion{S}{II}], and [\ion{O}{I}]. On the other hand, comparing [\ion{O}{III}] emission directly to broad H$\alpha$ emission \citep{Stern2012}, or X-ray luminosity \citep{Panessa2006}, describes a continuum of AGN as opposed to a bimodal distribution. 

In the case of AT2017bcc, its position on the BPT diagrams in Figure \ref{fig:bpt} indicate that its narrow line ratios are characteristic of a star-forming galaxy. It is likely that the observed narrow [\ion{O}{III}] emission did not purely arise from star formation, but any AGN contribution is presumably sub-dominant (otherwise the ratios would lie closer to the AGN region of the BPT diagram). 

We also measured the relative strength of its [\ion{O}{III}] lines to its broad H$\alpha$ line and X-ray emission. The relations in \cite{Stern2012} and \cite{Panessa2006} provide useful points of reference for these measurements, so we have compared our results to them in Figure \ref{fig:o3_emission}. It is clear from these metrics that AT2017bcc has much weaker narrow [\ion{O}{III}] emission than would be expected for a typical AGN: beyond $3\sigma$ for its H$\alpha$ luminosity, and beyond $1\sigma$ for its X-ray luminosity.

\subsection{Evidence for phase change}

Examples of AGN with weak or non-existent narrow emission lines have recently begun to emerge \citep{Greenwell2021}. These outliers may provide insight into transitional phases of AGN evolution. Two potential scenarios have been proposed to explain this transition.

The first is that the nucleus has been obscured by an influx of material, for example as a result of a recent galaxy merger. This cocoon would prevent the radiation emitted by accretion onto the SMBH from exciting the narrow line regions further out in the galaxy. Eventually, the new material causes the AGN to accrete at a higher rate, increasing the emitted radiation pressure and expelling the obscuring material. There is then a short epoch ($\sim 100$ yrs), as the newly unobscured light from the nucleus travels to the narrow line regions, when the AGN appears to have very weak [\ion{O}{III}] emission. In the case of a recent galaxy merger, we would expect to see signs in the host's morphology, and significant ongoing star formation caused by the material influx \citep{Goulding2009}. Our SED fits and radio observations show signs of ongoing star formation (though the radio more likely arises from previous episodes of AGN activity), and higher-resolution imaging could reveal traces of a recent merger. However, the very sudden brightening of the continuum in 2017, and unusual variability of the broad lines, are not naturally accounted for by a scenario of gradually infalling material obscuring the narrow lines, requiring an additional mechanism to fully explain our observations.

The second scenario is that the AGN activity has been recently triggered. This would not require a recent galaxy merger or a cocoon of obscuring material, but instead the serendipitous detection of a significant increase in the accretion rate. AT2017bcc showed negligible variation in luminosity for almost a decade of coverage, prior to the flare in 2017 and subsequent variability. It is well known that AGN go through phases of increased activity, thought to last on the order of $\sim 10^5$ yrs \citep{Schawinski2015}. The increase in luminosity and onset of significant variability in the last few years may mark the beginning of one of these active phases. The fact that the narrow [\ion{O}{III}] lines are weak also implies that the central SMBH had been quiescent for at least $\sim 100$ years (the approximate light travel time to the narrow-line region).

\subsection{Origin of initial flare}
Flares in galactic nuclei which defy classification are a relatively new class of object \citep{Kankare2017, Frederick2021}, and in the absence of a single physical explanation have been dubbed ambiguous nuclear transients (ANTs, \cite{Holoien2022}). Some ANTs have been attributed to gravitational microlensing \citep{Lawrence2016, Bruce2017}, but in the case of AT2017bcc the clear evolution in the spectral profile rules this out. Double-peaked emission features were also thought to arise from binary AGN \citep{Gaskell1983}. This hypothesis was refuted by \cite{Eracleous1997}, and \cite{Kelley2020} describes how double peaks would not be observable in such systems, so we do not consider it here.

We thus conclude that the initial flare most likely signals the beginning of an increase in the rate of accretion onto the central SMBH. This enhanced accretion rate is likely due to a rapid influx of material, either caused by disc instabilities or a disrupted star. As there may have been a dormant AGN disc prior to the flare in 2017, both of these mechanisms are plausible. Interestingly, the only other source for which we could identify a shark fin asymmetry in the central component of H$\alpha$ is ASASSN-14ko, a periodic nuclear transient in a galaxy hosting dual AGN, and a claimed repeating partial TDE \citep{Payne2021,Tucker2021}. Here, we explore the possibility that a TDE interacting with an existing accretion disk was the cause of the phase change in AT2017bcc.

Observationally, the flare bears some resemblance to a TDE. Its SED shows a strong blue continuum, with excess emission in the infra-red. This is seen in both TDEs and AGN flares \citep{Jiang2021}. The total radiated energy during the initial 2017 flare, $\sim {\rm few}\times10^{51}$\,erg, is also consistent with luminous optical TDEs. Spectral comparisons show that AT2017bcc's broad H$\alpha$ emission at early times is most similar to TDEs in the literature, while it develops more AGN-like features after the initial flare.  The broad component in ASASSN-14ko source could also be modelled with an accretion disc profile, suggesting AT2017bcc and ASASSN-14ko likely share a similar geometry in their line-forming regions. This may strengthen the case for a TDE in both objects, however this does not appear to repeat periodically in AT2017bcc (at least for periods to which we are sensitive, i.e.~$\lesssim$ a few years). The persistence of the broad line region in the spectra, years after the initial flare, indicates that there may have been a dormant AGN-like accretion disc and broad line region already present. There is a precedent for this kind of event; PS16dtm was a TDE observed in an already active AGN \citep{Blanchard2017}. In that case, X-ray emission from the pre-existing AGN disappeared during the flare, suggesting a disruption or obscuration of the AGN disc.

Fits to the host galaxy's SED show a small ($< 10\%$) AGN contribution to the IR luminosity. The host appears to contain an SMBH in the range of a few $\times 10^7 - 10^8\; {\rm M_{\rm \odot}}$, which straddles the Hills mass for a solar-type star. Thus a TDE origin would require either a rather massive star, or for the SMBH to exist in the lower ($< 10^8\; {\rm M_{\rm \odot}}$) part of this range, or to be rapidly rotating. While these conditions are not impossible, it is potentially more likely that other AGN flaring mechanisms are driving the recent activity. In particular, the bolometric luminosity $\sim10^{45}-10^{46}$\,erg\,s$^{-1}$ implied by the IR flare is challenging for TDE models. Therefore, even if a TDE was the initial trigger, the bulk of the luminosity produced since 2017 would likely arise from re-activation of a pre-existing disk, rather than accretion of the disrupted star.

\section{Conclusions}

We have conducted an extensive study of a nuclear transient, AT2017bcc, discovered by Pan-STARRS \citep{Chambers2017}. As it was found during the counterpart search for a GW signal, G274296 \citep{2017GCN.20689....1L}, we explored the possibility that this was a genuine multi-messenger event. Although we found the two signals to be likely unrelated, AT2017bcc was unique enough to study in its own right. We thus presented photometric follow-up in the radio, UV, optical, infra-red, and X-ray as well as optical spectroscopy.

Modelling archival SDSS magnitudes with \textsc{Prospector} showed the host to be a red galaxy with an old stellar population, and suggested the presence of a dormant AGN with a SMBH mass of a few $10^7-10^8\,{\rm M}_\odot$. Before it was discovered in 2017 the galaxy showed very little activity, varying in luminosity by $< 0.1$ mags in survey photometry. This historical quiescence was also shown in the weak [\ion{O}{III}] emission in recent spectra, which implied that the narrow-line emitting regions of the AGN had not been illuminated for at least $\sim 100$ years.

The flare which marked the discovery of AT2017bcc in 2017 was a long-lived nuclear transient with a strong blue continuum and broad H$\alpha$ emission. Since then, it has re-brightened multiple times on variable timescales. In some bands the subsequent peaks were greater in luminosity than the first. We calculated the structure function of this variability, and found it to be consistent with the observed quasar population \citep{Schmidt2010}. 

The broad spectral features resembled both TDEs \citep{Nicholl2020, Short2020} and double-peaked AGN \citep{Eracleous1994}, with an asymmetrical central component and a boxy broad component. These features showed evolution in their shape for years after the first luminosity peak. We modelled this spectral evolution using a circular accretion disc with a wind and a single spiral arm, and a double Gaussian representing a partially obscured outflow. This analysis suggested that the changing profiles were driven by a precession of the disc's inclination, the strength and location of the spiral arm, and the morphology of the outflow.

We conclude that the counterpart search in 2017 serendipitously caught the re-ignition of an AGN which had been dormant for at least a century. A plausible cause of the boosted accretion onto the SMBH is a TDE, or another dramatic event which appears to have disturbed the pre-existing accretion disc.

\section*{Acknowledgements}

ER is supported by the Science and Technology Facilities Council (STFC).

MN is supported by the European Research Council (ERC) under the European Union’s Horizon 2020 research and innovation programme (grant agreement No.~948381) and by UK Space Agency Grant No.~ST/Y000692/1.

MF is supported by a Royal Society - Science Foundation Ireland University Research Fellowship.

SM acknowledges support from the Research Council of Finland project 350458.

GP gratefully acknowledges support from a Royal Society University Research Fellowship URF{\textbackslash}R1{\textbackslash}221500 and RF{\textbackslash}ERE{\textbackslash}221015.  GP and PS acknowledge support from STFC grant ST/V005677/1.

TMR acknowledges the financial support of the Vilho, Yrj{\"o} and Kalle V{\"a}is{\"a}l{\"a} Foundation of the Finnish academy of Science and Letters.

{\L}W acknowledges funding from the European Union's Horizon 2020 research and innovation programme under grant agreement No. 101004719 (OPTICON-RadioNET Pilot, ORP).

JPA acknowledges funding from ANID, Millennium Science Initiative, ICN12\_009

T.-W.~Chen thanks the Max Planck Institute for Astrophysics for hosting her as a guest researcher.

CPG acknowledges financial support from the Secretary of Universities
and Research (Government of Catalonia) and by the Horizon 2020 Research
and Innovation Programme of the European Union under the Marie
Sk\l{}odowska-Curie and the Beatriu de Pin\'os 2021 BP 00168 programme,
from the Spanish Ministerio de Ciencia e Innovaci\'on (MCIN) and the
Agencia Estatal de Investigaci\'on (AEI) 10.13039/501100011033 under the
PID2020-115253GA-I00 HOSTFLOWS project, and the program Unidad de
Excelencia Mar\'ia de Maeztu CEX2020-001058-M.

G.L. is supported by a research grant (19054) from VILLUM FONDEN.

The Pan-STARRS telescopes are supported by NASA Grants NNX12AR65G and NNX14AM74G. 

Based on observations collected at the European Organisation for Astronomical Research in the Southern Hemisphere, Chile, as part of ePESSTO/ePESSTO+ (the extended/advanced Public ESO Spectroscopic Survey for Transient Objects Survey).

ATLAS is primarily funded through NASA grants NN12AR55G, 80NSSC18K0284, and 80NSSC18K1575. The ATLAS science products are provided by the University of Hawaii, QUB, STScI, SAAO and Millennium Institute of Astrophysics in Chile. 

This work is based in part on observations obtained at the MDM Observatory, operated by Dartmouth College, Columbia University, Ohio State University, Ohio University, and the University of Michigan.

This research has made use of data or software obtained from the Gravitational Wave Open Science Center (gwosc.org), a service of the LIGO Scientific Collaboration, the Virgo Collaboration, and KAGRA. This material is based upon work supported by NSF's LIGO Laboratory which is a major facility fully funded by the National Science Foundation, as well as the Science and Technology Facilities Council (STFC) of the United Kingdom, the Max-Planck-Society (MPS), and the State of Niedersachsen/Germany for support of the construction of Advanced LIGO and construction and operation of the GEO600 detector. Additional support for Advanced LIGO was provided by the Australian Research Council. Virgo is funded, through the European Gravitational Observatory (EGO), by the French Centre National de Recherche Scientifique (CNRS), the Italian Istituto Nazionale di Fisica Nucleare (INFN) and the Dutch Nikhef, with contributions by institutions from Belgium, Germany, Greece, Hungary, Ireland, Japan, Monaco, Poland, Portugal, Spain. KAGRA is supported by Ministry of Education, Culture, Sports, Science and Technology (MEXT), Japan Society for the Promotion of Science (JSPS) in Japan; National Research Foundation (NRF) and Ministry of Science and ICT (MSIT) in Korea; Academia Sinica (AS) and National Science and Technology Council (NSTC) in Taiwan.

Computations were performed using resources provided by Supercomputing Wales, funded by STFC grants ST/I006285/1 and ST/V001167/1 supporting the UK Involvement in the Operation of Advanced LIGO.

This research has made use of the CIRADA cutout service at URL cutouts.cirada.ca, operated by the Canadian Initiative for Radio Astronomy Data Analysis (CIRADA). CIRADA is funded by a grant from the Canada Foundation for Innovation 2017 Innovation Fund (Project 35999), as well as by the Provinces of Ontario, British Columbia, Alberta, Manitoba and Quebec, in collaboration with the National Research Council of Canada, the US National Radio Astronomy Observatory and Australia’s Commonwealth Scientific and Industrial Research Organisation.

\section*{Data availability}

All data in this paper will be made publicly available via WISeREP \citep{Yaron2012}.

\bibliographystyle{mnras}
\bibliography{main}

\section*{Affiliations}

\textit{$^1$School of Physics and Astronomy, University of Birmingham, Edgbaston, B15 2TT, England\\
$^2$Institute for Gravitational Wave Astronomy, University of Birmingham, Edgbaston, B15 2TT, England\\
$^3$Astrophysics Research Centre, School of Mathematics and Physics, Queen’s University Belfast, BT7 1NN, UK\\
$^4$Department of Astrophysical Sciences, Princeton University, Princeton, NJ 08544, USA\\
$^5$Center for Interdisciplinary Exploration and Research in Astrophysics
(CIERA), Northwestern University, 1800 Sherman Ave. 8th Floor,
Evanston, IL 60201, USA\\
$^6$Department of Astronomy, University of California, Berkeley, CA 94720-3411, USA\\
$^7$School of Physics, O'Brien Centre for Science North, University College Dublin, Belfield, Dublin 4, Ireland\\
$^8$Space Telescope Science Institute, 3700 San Martin Dr, Baltimore, MD 21218, USA\\
$^9$Department of Physics and Astronomy, University of Turku, 20500 Turku, Finland\\
$^{10}$School of Sciences, European University Cyprus, Diogenes Street, Engomi, 1516, Nicosia, Cyprus\\
$^{11}$Department of Physics and Astronomy, Vanderbilt University, Nashville, TN 37235\\
$^{12}$Department of Astronomy/Steward Observatory, 933 North Cherry Avenue, Rm. N204, Tucson, AZ 85721-0065, USA\\
$^{13}$Astronomical Observatory, University of Warsaw, Al. Ujazdowskie 4, 00-478 Warszawa, Poland\\
$^{14}$Institute for Astronomy, University of Edinburgh, Royal Observatory, Blackford Hill, Edinburgh EH9 3HJ, UK\\
$^{15}$Cosmic Dawn Center (DAWN), Denmark\\
$^{16}$Niels Bohr Institute, University of Copenhagen, Jagtvej 128, 2200
Copenhagen N, Denmark.\\
$^{17}$European Southern Observatory, Alonso de Córdova 3107, Casilla 19, Santiago, Chile\\
$^{18}$Millennium Institute of Astrophysics MAS, Nuncio Monsenor Sotero Sanz 100, Off.
104, Providencia, Santiago, Chile\\
$^{19}$INAF-Osservatorio Astronomico di Padova, Vicolo dell’Osservatorio 5, I-35122 Padova, Italy\\
$^{20}$Center for Astrophysics, Harvard \& Smithsonian, 60 Garden Street, Cambridge, MA 02138-1516, USA\\
$^{21}$Institute for Astronomy, University of Hawai’i, 2680 Woodlawn Drive, Honolulu, HI 96822, USA\\
$^{22}$Technische Universit{\"a}t M{\"u}nchen, TUM School of Natural Sciences, Physik-Department, James-Franck-Stra{\ss}e 1, 85748 Garching, Germany\\
$^{23}$Institut d’Estudis Espacials de Catalunya (IEEC), Gran Capit\`a, 2-4,
Edifici Nexus, Desp. 201, E-08034 Barcelona, Spain \\
$^{24}$Institute of Space Sciences (ICE, CSIC), Campus UAB, Carrer de Can
Magrans, s/n, E-08193 Barcelona, Spain \\
$^{25}$Cardiff Hub for Astrophysics Research and Technology, School of Physics \& Astronomy, Cardiff University, Queens Buildings, The Parade, Cardiff, CF24 3AA, UK\\
$^{26}$Finnish Centre for Astronomy with ESO (FINCA), University of Turku, 20014 Turku, Finland\\
$^{27}$Tuorla Observatory, Department of Physics and Astronomy, University of Turku, 20014 Turku, Finland\\
$^{28}$DTU Space, National Space Institute, Technical University of Denmark, Elektrovej 327, 2800 Kgs. Lyngby, Denmark\\
$^{29}$The School of Physics and Astronomy, Tel Aviv University, Tel Aviv 69978, Israel\\
$^{30}$Department of Physics, University of Oxford, Denys Wilkinson Building, Keble Road, Oxford OX1 3RH, UK\\
$^{31}$Department of Astronomy, University of Illinois at Urbana-Champaign, 1002 W. Green Street, Urbana, IL 61801, USA\\
$^{32}$Sternberg Astronomical Institute, Moscow State University, Universitetskii pr. 13, 119992 Moscow, Russia}

\appendix

\section{Additional plots and tables}\label{sec:appendix}

\begin{figure*}
    \centering
    \includegraphics[width=\linewidth, trim=0cm 0cm 0cm 2cm, clip]{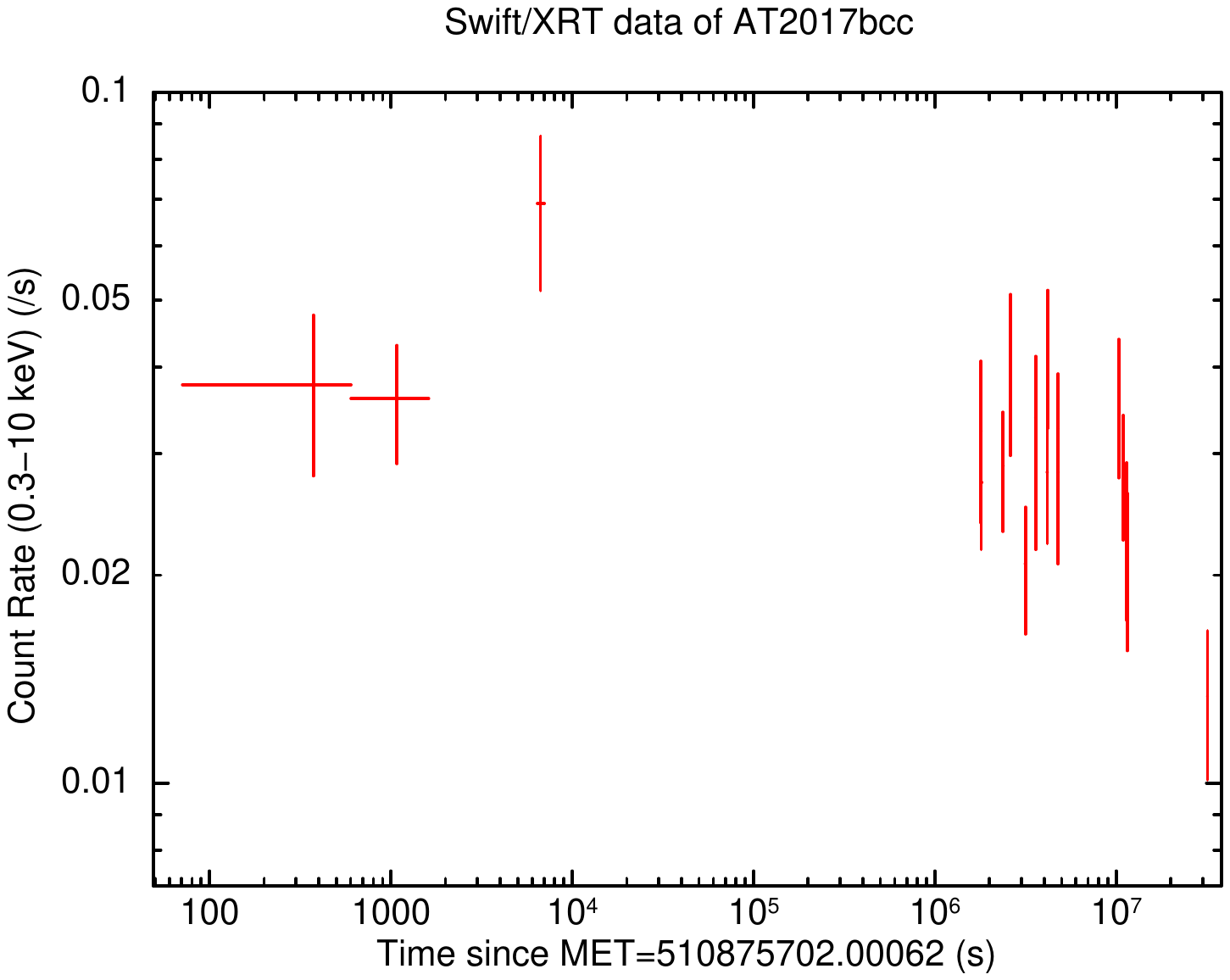}
    \caption{Swift/XRT light curve for AT2017bcc, also shown in Table \ref{tab:xray}.}
\end{figure*}

\begin{table*}
    \centering
    \caption{X-ray light curve for AT2017bcc. The conversion from counts to flux (unabsorbed, 0.3-10\,keV) for the best-fit power-law model is $4.21\times10^{-11}$\,erg\,cm$^{-2}$\,ct$^{-1}$.}
    \label{tab:xray}
    \bgroup
    \def\arraystretch{1.5}
    \begin{tabular}{cc}
        \hline
        MJD&counts [sec$^{-1}$]\\
        \hline
        $57822.917$&$0.03120^{+0.01125}_{-0.00909}$\\
        $57822.921$&$0.03858^{+0.01384}_{-0.01118}$\\
        $57822.925$&$0.04309^{+0.01532}_{-0.01239}$\\
        $57822.928$&$0.03553^{+0.01150}_{-0.00947}$\\
        $57822.990$&$0.04266^{+0.01111}_{-0.01111}$\\
        $57843.703$&$0.04142^{+0.01485}_{-0.01200}$\\
        $57843.708$&$0.03251^{+0.01192}_{-0.00963}$\\
        $57843.714$&$0.02923^{+0.00721}_{-0.00721}$\\
        $57850.483$&$0.02779^{+0.00830}_{-0.00693}$\\
        $57853.145$&$0.04136^{+0.01084}_{-0.01084}$\\
        $57859.458$&$0.02458^{+0.00573}_{-0.00573}$\\
        $57864.504$&$0.04424^{+0.01584}_{-0.01281}$\\
        $57864.510$&$0.02696^{+0.00916}_{-0.00748}$\\
        $57864.570$&$0.03244^{+0.01163}_{-0.00940}$\\
        $57864.576$&$0.02479^{+0.00772}_{-0.00641}$\\
        $57871.080$&$0.02881^{+0.00858}_{-0.00717}$\\
        $57871.415$&$0.06136^{+0.02192}_{-0.01771}$\\
        $57871.419$&$0.03210^{+0.01153}_{-0.00932}$\\
        $57878.319$&$0.03748^{+0.01348}_{-0.01089}$\\
        $57878.324$&$0.03795^{+0.01366}_{-0.01104}$\\
        $57878.330$&$0.02818^{+0.00926}_{-0.00774}$\\
        $57942.427$&$0.02704^{+0.00984}_{-0.00795}$\\
        $57942.432$&$0.04365^{+0.01066}_{-0.01066}$\\
        $57942.493$&$0.02134^{+0.00732}_{-0.00598}$\\
        $57942.498$&$0.05483^{+0.01342}_{-0.01342}$\\
        $57949.618$&$0.02670^{+0.00662}_{-0.00662}$\\
        $57955.988$&$0.02106^{+0.00540}_{-0.00540}$\\
        $58191.653$&$0.01857^{+0.00581}_{-0.00482}$\\
        \hline
    \end{tabular}
    \egroup
\end{table*}

\begin{figure*}
    \centering
    \includegraphics[width=\linewidth, trim=0cm 0cm 0cm 2cm, clip]{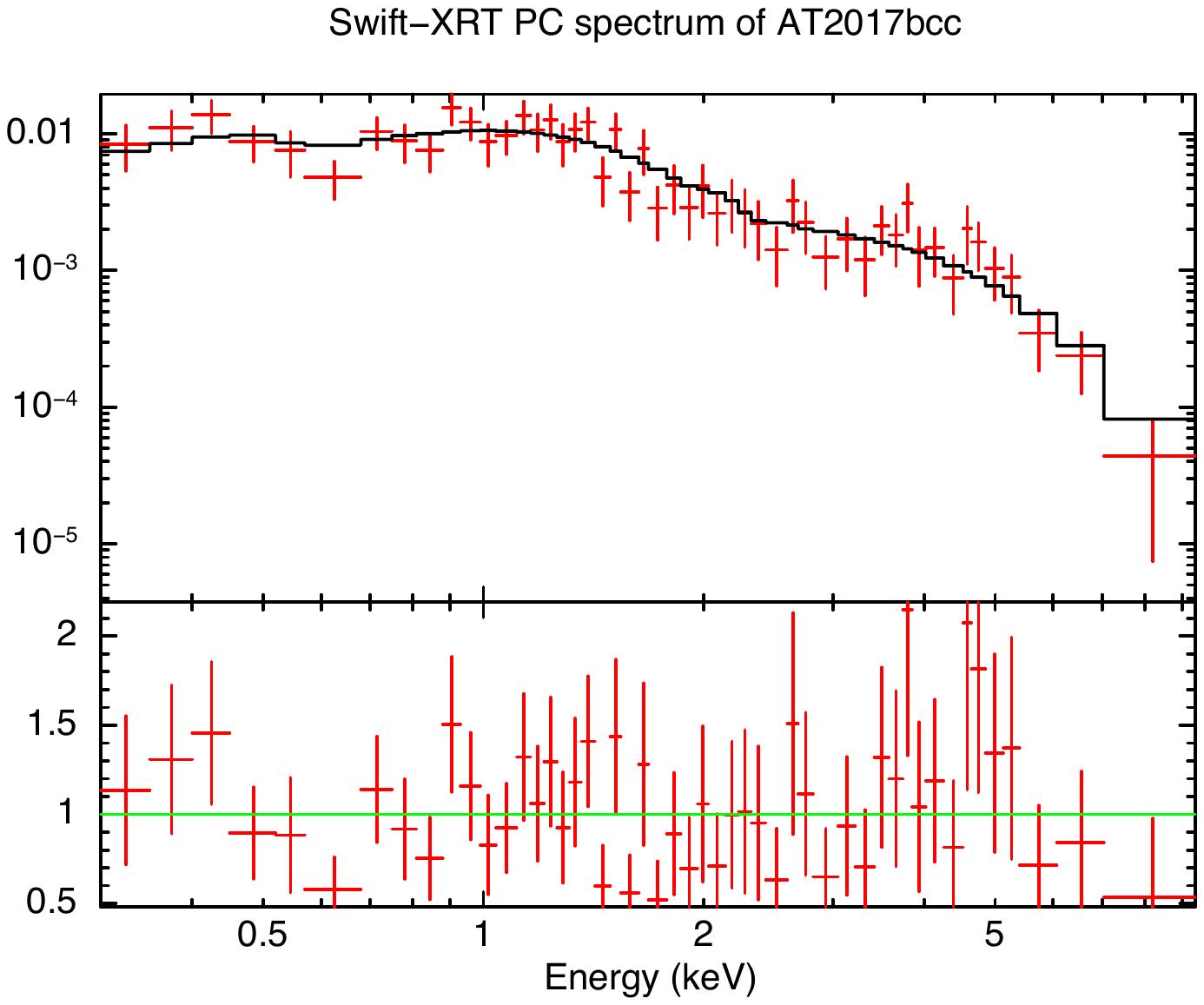}
    \caption{Swift-XRT PC spectrum of AT2017bcc.}
\end{figure*}

\begin{table*}
    \centering
    \caption{Radio observations of AT2017bcc.}
    \label{tab:radio}
    \bgroup
    \def\arraystretch{1.5}
    \begin{tabular}{cccc}
        \hline
        MJD&freq (GHz)&F ($\mu$Jy)&dF ($\mu$Jy)\\
        \hline
        $57833$&$4.999$&$354.2$&$22.9$\\
        $57833$&$7.099$&$358.2$&$18.1$\\
        $57833$&$19.199$&$173.6$&$41.7$\\
        $57833$&$24.499$&$140.5$&$48.0$\\
        $57835$&$2.679$&$373.8$&$46.8$\\
        $57835$&$3.523$&$366.6$&$33.4$\\
        $57835$&$4.999$&$422.8$&$25.5$\\
        $57835$&$7.099$&$378.5$&$16.3$\\
        $57835$&$8.549$&$366.5$&$17.0$\\
        $57835$&$10.999$&$314.5$&$13.6$\\
        $57835$&$13.499$&$256.7$&$12.3$\\
        $57835$&$15.999$&$222.9$&$15.0$\\
        $57869$&$2.679$&$212.1$&$74.1$\\
        $57869$&$3.523$&$309.6$&$30.8$\\
        $57869$&$4.999$&$357.7$&$35.5$\\
        $57869$&$7.099$&$340.7$&$25.3$\\
        $57869$&$8.549$&$326.8$&$19.5$\\
        $57869$&$10.999$&$305.1$&$15.2$\\
        $57869$&$13.499$&$282.4$&$13.5$\\
        $57869$&$15.999$&$271.4$&$16.7$\\
        $57934$&$2.679$&$375.1$&$35.3$\\
        $57934$&$3.523$&$375.5$&$23.8$\\
        $57934$&$4.999$&$350.7$&$23.0$\\
        $57934$&$7.099$&$307.5$&$24.5$\\
        $57934$&$8.549$&$351.3$&$24.1$\\
        $57934$&$10.999$&$322.0$&$19.5$\\
        $57934$&$13.499$&$325.2$&$33.2$\\
        $57934$&$15.999$&$321.9$&$36.3$\\
        $58135$&$2.679$&$332.1$&$26.0$\\
        $58135$&$3.523$&$345.2$&$13.3$\\
        $58135$&$4.999$&$372.9$&$26.1$\\
        $58135$&$7.099$&$336.8$&$20.1$\\
        $58135$&$8.549$&$311.7$&$19.5$\\
        $58135$&$10.999$&$269.1$&$17.4$\\
        $58135$&$13.499$&$286.1$&$18.8$\\
        $58135$&$15.999$&$295.0$&$19.6$\\
        \hline
    \end{tabular}
    \egroup
\end{table*}

\begin{figure*}
    \centering
    \includegraphics[width=\linewidth]{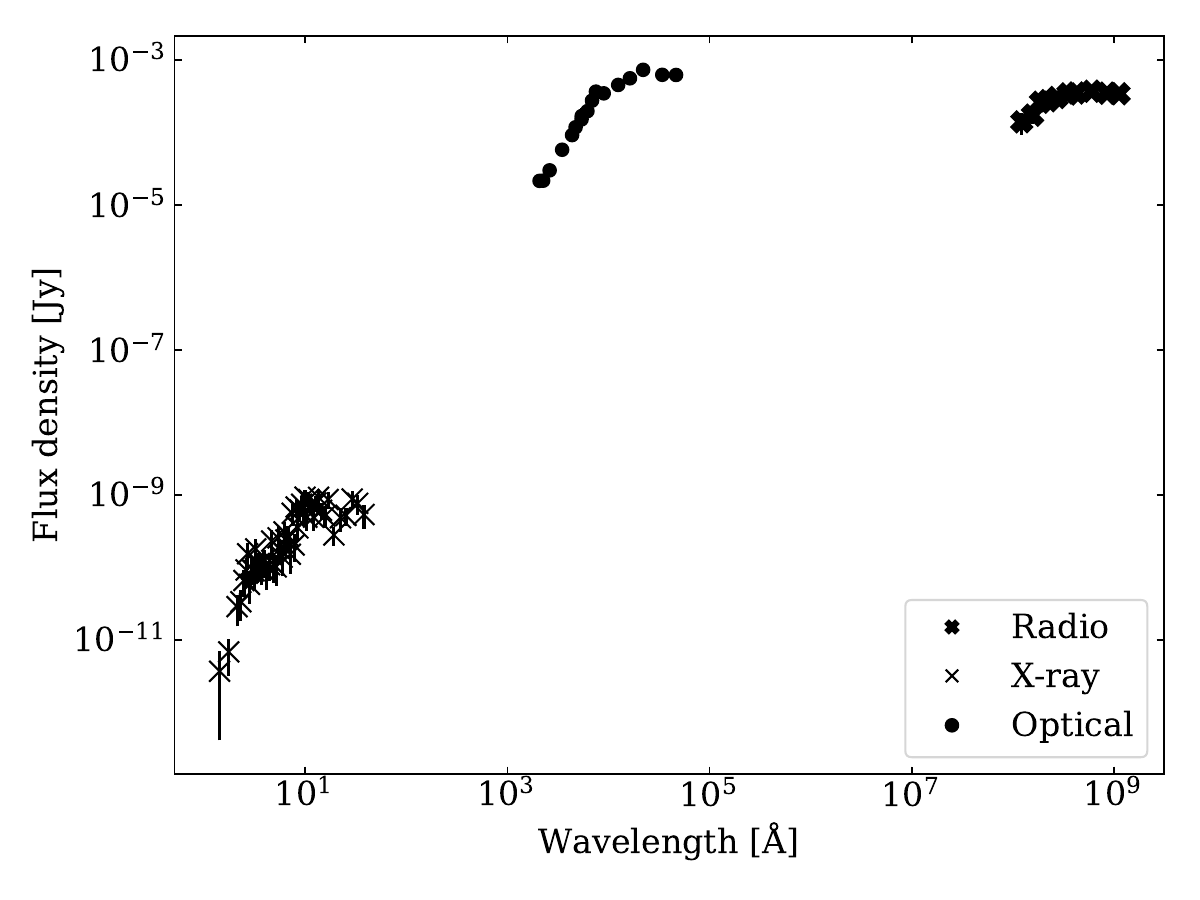}
    \caption{Spectral energy distribution across all observed wavelengths of AT2017bcc. The optical fluxes included are from the 2017-05 epoch shown in Figure \ref{fig:sed}. The X-ray and radio observations are described in Sections \ref{sec:xray} and \ref{sec:radio} respectively.}
\end{figure*}

\begin{figure*}
    \centering
    \includegraphics[width=\linewidth]{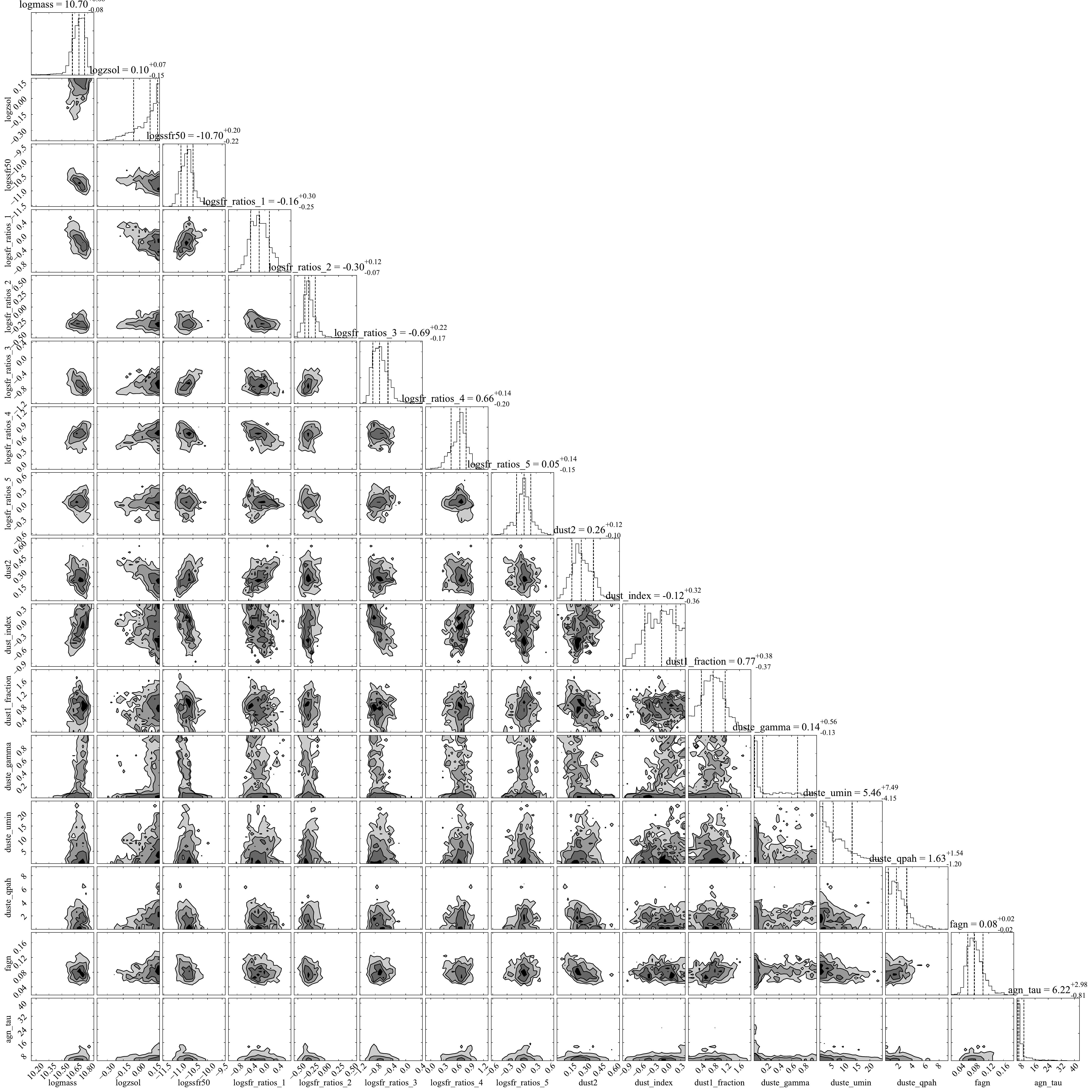}
    \caption{Corner plot showing results of fitting to host photometry of AT2017bcc using \textsc{Prospector}.}
\end{figure*}

\begin{figure*}
    \centering
    \includegraphics[width=\linewidth, trim=0cm 1cm 0cm 0cm]{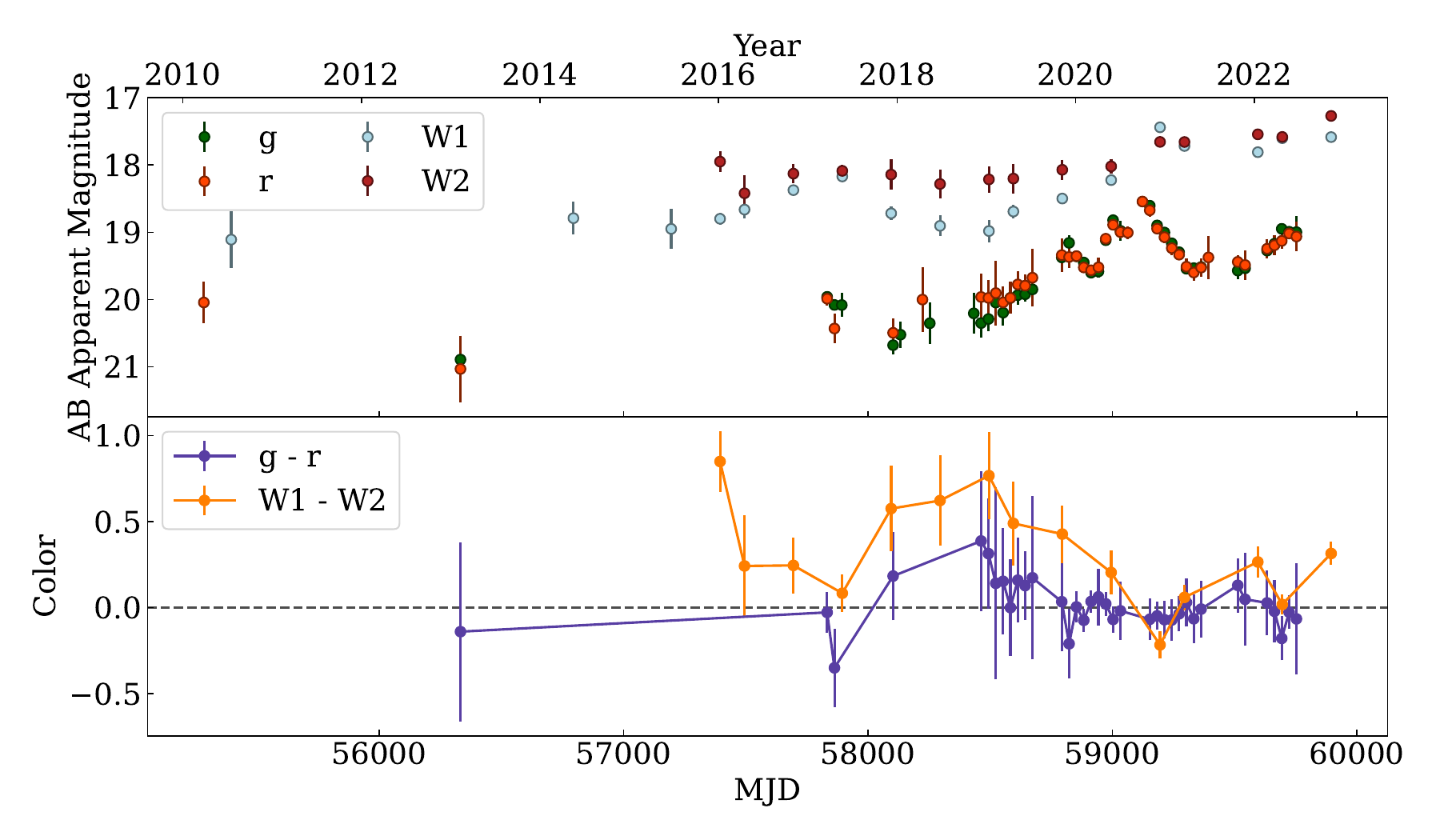}
    \caption{Host-subtracted light curve of AT2017bcc, with the optical and infra-red colour evolution plotted below. Each point is the result of binning fluxes to a monthly cadence before converting to magnitudes and calculating colours. All magnitudes are corrected for galactic extinction and host-subtracted using inferred host fluxes from fits to archival SDSS, 2MASS, and WISE photometry, described in Section \ref{sec:host}. The optical colours are largely consistent with zero, and are too noisy to show evolution. There may be some evolution in the mid infra-red.}
    \label{fig:at2017bcc_hslc}
\end{figure*}

\begin{table*}
	\centering
	\caption{Best-fit disc parameters from modelling the H$\alpha$ and H$\beta$ broad-line regions with the circular accretion disc model from \citet{Chen1989}. We show the following H$\alpha$ disc parameters: inner radius $\xi_{1}$, outer radius $\xi_{2}$, turbulent broadening $\sigma$ (km/s), inclination angle $i$ (deg), emissivity power law index $q$, spiral arm amplitude expressed as contrast ratio $A_{\rm s}$, spiral arm phase $\phi$ (deg), spiral arm pitch angle $\psi$ (deg), spiral arm width $w$ (deg), wind opening angle $\theta$ (deg), Log wind amplitude $t_0$ and wind optical depth $\tau$.}
	\label{tab:discparams}
        \bgroup
        \def\arraystretch{1.5}
        \rotatebox{270}{
	\begin{tabular}{lcccccccccccc} 
		\hline
		Epoch&$\xi_{1}$&$\xi_{2}$&$\sigma$ (km/s) &$i$ (deg)&$q$&$A_{\rm s}$&$\phi$ (deg)&$\psi$ (deg)&$w$ (deg)&$\theta$ (deg)&Log $t_0$&$\tau$\\\\
		\hline
2017-02-19&$700.6^{+18.5}_{-20.3}$&$4.8^{+5.1}_{-7.5}$&$1267.8^{+1.5}_{-1.3}$&$89.3^{+34.0}_{-0.6}$&$1.8^{+1.2}_{-0.8}$&$3.3^{+2.0}_{-2.0}$&$42.5^{+47.5}_{-76.4}$&$40.2^{+43.2}_{-139.6}$&$51.6^{+20.2}_{-22.7}$&$0.7^{+0.4}_{-0.4}$&$-1.2^{+17.7}_{-12.8}$&$0.9^{+0.8}_{-0.9}$\\
2017-02-27&$689.2^{+33.7}_{-25.5}$&$14.8^{+0.9}_{-0.9}$&$1266.2^{+1.1}_{-1.5}$&$75.8^{+4.5}_{-6.3}$&$2.0^{+0.6}_{-0.5}$&$2.7^{+0.4}_{-0.5}$&$47.2^{+13.7}_{-17.3}$&$87.8^{+15.0}_{-5.0}$&$80.4^{+24.1}_{-19.4}$&$0.3^{+0.7}_{-0.7}$&$-15.2^{+6.5}_{-8.2}$&$0.7^{+0.3}_{-0.5}$\\
2017-03-20&$723.4^{+34.6}_{-19.0}$&$10.8^{+12.4}_{-3.2}$&$1267.2^{+1.8}_{-0.8}$&$79.8^{+10.7}_{-29.3}$&$1.9^{+0.2}_{-0.3}$&$2.0^{+0.5}_{-1.2}$&$46.8^{+15.7}_{-91.2}$&$79.9^{+152.6}_{-18.4}$&$59.7^{+39.6}_{-9.9}$&$0.2^{+0.3}_{-1.2}$&$-4.8^{+11.4}_{-13.0}$&$0.5^{+0.1}_{-1.1}$\\
2017-03-24&$701.0^{+27.9}_{-36.1}$&$2.2^{+1.0}_{-18.6}$&$1267.3^{+0.9}_{-1.3}$&$89.7^{+12.2}_{-0.4}$&$2.1^{+1.0}_{-0.4}$&$2.7^{+2.6}_{-2.5}$&$20.1^{+21.8}_{-46.2}$&$64.8^{+71.5}_{-32.7}$&$72.2^{+38.6}_{-33.6}$&$0.7^{+0.8}_{-1.5}$&$-0.4^{+15.4}_{-11.0}$&$0.6^{+0.4}_{-1.1}$\\
2017-04-04&$697.6^{+20.0}_{-19.9}$&$6.9^{+5.5}_{-5.2}$&$1267.8^{+2.1}_{-3.2}$&$86.7^{+5.2}_{-8.1}$&$2.0^{+0.7}_{-0.5}$&$2.7^{+2.1}_{-4.1}$&$27.9^{+34.9}_{-40.7}$&$75.3^{+79.8}_{-42.0}$&$57.2^{+58.4}_{-23.9}$&$0.7^{+0.3}_{-0.2}$&$-5.6^{+9.2}_{-22.8}$&$1.2^{+0.6}_{-0.8}$\\
2017-04-23&$700.4^{+37.5}_{-37.6}$&$2.8^{+1.7}_{-5.8}$&$1267.5^{+2.0}_{-1.7}$&$89.3^{+2.8}_{-1.2}$&$1.7^{+0.9}_{-1.0}$&$2.5^{+0.7}_{-1.0}$&$28.7^{+36.3}_{-59.3}$&$32.2^{+21.8}_{-23.7}$&$53.3^{+26.0}_{-18.0}$&$0.8^{+1.5}_{-3.2}$&$0.0^{+15.2}_{-10.5}$&$0.7^{+0.5}_{-1.6}$\\
2017-06-29&$704.9^{+30.9}_{-19.2}$&$2.4^{+1.6}_{-10.8}$&$1267.7^{+2.3}_{-2.7}$&$89.4^{+4.0}_{-0.5}$&$1.4^{+1.8}_{-1.6}$&$2.3^{+1.4}_{-0.9}$&$27.3^{+47.4}_{-56.4}$&$34.2^{+47.3}_{-98.4}$&$65.6^{+38.4}_{-55.9}$&$0.7^{+0.3}_{-0.3}$&$-3.7^{+11.1}_{-16.5}$&$1.0^{+1.1}_{-1.1}$\\
2017-11-26&$700.5^{+23.1}_{-25.2}$&$1.5^{+0.9}_{-1.2}$&$1267.7^{+1.0}_{-1.0}$&$80.2^{+30.8}_{-20.7}$&$1.4^{+1.8}_{-1.7}$&$3.2^{+2.2}_{-1.9}$&$22.7^{+23.8}_{-32.6}$&$26.9^{+24.9}_{-139.2}$&$59.9^{+25.4}_{-18.8}$&$1.0^{+0.7}_{-0.6}$&$-1.9^{+9.2}_{-14.9}$&$1.0^{+0.8}_{-1.1}$\\
2017-12-29&$698.0^{+27.8}_{-21.9}$&$1.2^{+0.3}_{-0.4}$&$1268.2^{+1.4}_{-1.5}$&$82.4^{+15.2}_{-9.5}$&$1.3^{+1.5}_{-1.2}$&$2.5^{+0.8}_{-0.9}$&$19.6^{+18.3}_{-22.7}$&$29.1^{+31.0}_{-38.9}$&$47.7^{+25.9}_{-28.6}$&$0.8^{+1.1}_{-1.2}$&$-3.0^{+6.2}_{-14.1}$&$1.1^{+0.8}_{-0.8}$\\
2018-01-05&$707.5^{+23.6}_{-15.1}$&$2.0^{+0.9}_{-2.4}$&$1267.5^{+2.4}_{-1.8}$&$82.0^{+13.2}_{-9.7}$&$1.6^{+1.5}_{-1.2}$&$2.7^{+1.8}_{-1.4}$&$41.1^{+40.7}_{-37.1}$&$20.7^{+28.4}_{-68.0}$&$59.7^{+14.1}_{-11.9}$&$0.7^{+0.9}_{-0.9}$&$-5.7^{+5.3}_{-4.7}$&$0.9^{+0.5}_{-0.8}$\\
2018-03-17&$690.8^{+36.7}_{-33.7}$&$2.7^{+0.8}_{-2.8}$&$1267.7^{+1.6}_{-0.4}$&$78.7^{+10.5}_{-15.3}$&$1.8^{+0.7}_{-0.1}$&$3.9^{+1.3}_{-0.5}$&$77.8^{+26.6}_{-28.7}$&$-27.2^{+72.1}_{-95.5}$&$13.6^{+8.3}_{-65.8}$&$1.0^{+0.3}_{-0.2}$&$-4.7^{+1.2}_{-4.0}$&$0.5^{+0.1}_{-0.2}$\\
2018-05-15&$703.6^{+46.8}_{-30.9}$&$4.6^{+0.8}_{-0.8}$&$1267.7^{+1.1}_{-1.1}$&$83.3^{+11.9}_{-7.6}$&$0.8^{+1.0}_{-1.8}$&$3.5^{+1.2}_{-1.8}$&$122.0^{+14.3}_{-15.2}$&$25.0^{+5.5}_{-4.9}$&$14.5^{+7.1}_{-5.4}$&$0.1^{+0.1}_{-0.5}$&$-6.2^{+1.5}_{-1.9}$&$0.6^{+0.1}_{-0.3}$\\
2019-01-26&$697.5^{+19.3}_{-16.5}$&$4.0^{+1.7}_{-3.4}$&$1267.0^{+1.8}_{-1.4}$&$84.3^{+14.1}_{-49.7}$&$1.5^{+2.0}_{-1.4}$&$3.6^{+2.6}_{-3.3}$&$84.5^{+33.8}_{-32.9}$&$-56.4^{+76.4}_{-417.8}$&$41.2^{+29.2}_{-55.0}$&$1.0^{+0.9}_{-1.0}$&$-4.7^{+8.7}_{-13.5}$&$0.6^{+0.3}_{-0.8}$\\
2020-01-28&$694.9^{+26.3}_{-21.7}$&$4.6^{+0.4}_{-0.4}$&$1280.3^{+6.5}_{-6.5}$&$83.2^{+3.6}_{-3.0}$&$1.7^{+0.6}_{-0.6}$&$2.6^{+0.4}_{-0.6}$&$87.8^{+17.3}_{-14.4}$&$18.5^{+3.6}_{-3.8}$&$58.7^{+7.9}_{-9.2}$&$0.8^{+0.3}_{-0.3}$&$-9.2^{+1.1}_{-1.2}$&$0.5^{+0.1}_{-0.1}$\\
2021-01-21&$699.9^{+25.4}_{-26.1}$&$5.9^{+3.3}_{-6.0}$&$1267.1^{+3.4}_{-3.5}$&$79.9^{+9.5}_{-9.2}$&$1.7^{+0.9}_{-1.0}$&$3.5^{+3.4}_{-3.2}$&$85.3^{+53.3}_{-43.4}$&$23.0^{+10.8}_{-37.5}$&$39.1^{+28.3}_{-29.1}$&$0.9^{+0.7}_{-0.7}$&$-3.2^{+9.6}_{-15.5}$&$0.6^{+0.2}_{-0.4}$\\
2022-02-07&$708.0^{+52.0}_{-134.4}$&$5.4^{+1.6}_{-1.1}$&$1264.9^{+2.0}_{-1.7}$&$72.1^{+4.8}_{-14.8}$&$2.0^{+0.6}_{-0.3}$&$6.3^{+1.6}_{-2.1}$&$100.1^{+35.5}_{-33.8}$&$16.9^{+2.0}_{-1.7}$&$54.4^{+8.3}_{-18.6}$&$1.2^{+0.4}_{-0.4}$&$-5.8^{+1.6}_{-12.3}$&$0.5^{+0.1}_{-0.1}$\\
2023-01-15&$604.7^{+52.0}_{-67.2}$&$8.0^{+1.7}_{-2.1}$&$1264.0^{+5.6}_{-6.6}$&$45.4^{+2.4}_{-2.6}$&$1.4^{+2.0}_{-1.6}$&$11.6^{+1.0}_{-0.9}$&$160.8^{+28.0}_{-37.3}$&$12.4^{+1.4}_{-1.4}$&$88.2^{+11.1}_{-4.1}$&$0.2^{+0.3}_{-0.6}$&$-6.9^{+3.6}_{-3.2}$&$0.7^{+0.5}_{-0.5}$\\
		\hline
	\end{tabular}}
        \egroup
\end{table*}

\begin{table*}
	\centering
	\caption{Best-fit parameters for the two Gaussian model for the central shark fin structure. We show the width $\sigma_b$(\AA) and $\sigma_{b2}$(\AA) for each component and the offset in from the central rest H$\alpha$ wavelength $\Delta_1$(\AA) and $\Delta_{2}$(\AA)}
	\label{tab:outflowparams}
        \bgroup
        \def\arraystretch{1.5}
	\begin{tabular}{lcccc} 
		\hline
		Epoch&$\sigma_b$(\AA)&$\sigma_{b2}$(\AA)&$\Delta_1$(\AA) &$\Delta_{2}$(\AA)\\
		\hline
2017-02-19&$64.5^{+19.6}_{-23.0}$&$35.7^{+20.5}_{-21.9}$&$-23.7^{+3.7}_{-4.5}$&$-51.9^{+20.2}_{-16.5}$\\
2017-02-27&$43.2^{+4.9}_{-5.5}$&$15.5^{+5.4}_{-6.4}$&$-17.7^{+5.3}_{-6.0}$&$-44.1^{+12.4}_{-13.9}$\\
2017-03-20&$68.4^{+9.6}_{-9.9}$&$25.4^{+13.2}_{-12.3}$&$-22.2^{+1.7}_{-1.7}$&$-53.9^{+10.3}_{-10.3}$\\
2017-03-24&$71.0^{+28.5}_{-42.2}$&$39.5^{+15.3}_{-45.0}$&$-26.4^{+2.8}_{-2.3}$&$-45.6^{+25.1}_{-56.2}$\\
2017-04-04&$63.1^{+20.8}_{-17.7}$&$33.3^{+7.7}_{-8.8}$&$-20.9^{+1.6}_{-1.7}$&$-56.4^{+8.8}_{-16.4}$\\
2017-04-23&$79.9^{+57.3}_{-62.4}$&$27.5^{+5.5}_{-7.1}$&$-19.7^{+3.6}_{-3.4}$&$-38.4^{+22.7}_{-13.5}$\\
2017-06-29&$62.4^{+28.1}_{-25.4}$&$27.5^{+5.3}_{-4.8}$&$-17.3^{+2.3}_{-2.0}$&$-28.1^{+4.2}_{-4.4}$\\
2017-11-26&$54.4^{+29.6}_{-32.3}$&$38.3^{+19.8}_{-33.6}$&$-23.4^{+5.2}_{-8.0}$&$-42.0^{+21.2}_{-18.8}$\\
2017-12-29&$78.5^{+24.2}_{-26.1}$&$52.1^{+23.2}_{-23.9}$&$-20.3^{+5.3}_{-5.1}$&$-44.1^{+20.4}_{-16.9}$\\
2018-01-05&$56.5^{+32.8}_{-28.9}$&$28.8^{+7.1}_{-13.0}$&$-16.8^{+10.9}_{-7.0}$&$-36.1^{+19.8}_{-10.0}$\\
2018-03-17&$60.8^{+61.5}_{-37.9}$&$26.3^{+71.5}_{-2.7}$&$-18.7^{+21.7}_{-13.0}$&$-26.6^{+58.6}_{-6.7}$\\
2018-05-15&$126.3^{+25.6}_{-17.7}$&$8.7^{+2.5}_{-2.5}$&$-28.7^{+3.5}_{-2.5}$&$-13.1^{+1.5}_{-2.3}$\\
2019-01-26&$79.9^{+24.6}_{-31.1}$&$39.2^{+9.9}_{-17.5}$&$-20.8^{+9.0}_{-7.6}$&$-51.2^{+16.7}_{-19.3}$\\
2020-01-28&$56.6^{+3.5}_{-3.5}$&$38.5^{+1.9}_{-1.5}$&$-16.0^{+0.9}_{-0.8}$&$-40.9^{+2.5}_{-2.3}$\\
2021-01-21&$70.8^{+27.2}_{-28.6}$&$32.0^{+12.3}_{-13.9}$&$-19.5^{+3.0}_{-2.2}$&$-42.0^{+29.2}_{-16.3}$\\
2022-02-07&$102.6^{+88.6}_{-90.2}$&$22.9^{+1.7}_{-1.8}$&$-34.6^{+17.1}_{-42.2}$&$-21.5^{+1.0}_{-1.4}$\\
2023-01-15&$169.4^{+15.5}_{-14.5}$&$3.2^{+3.3}_{-2.6}$&$-18.2^{+1.4}_{-1.6}$&$-51.0^{+2.9}_{-1.7}$\\
		\hline
	\end{tabular}
        \egroup
\end{table*}

\begin{table*}
	\centering
	\caption{Best-fit parameters for the narrow emission line flux ratios based on modelling of the high S/N 2020-01-28 spectrum.}
	\label{tab:narrowlinefluxes}
        \bgroup
        \def\arraystretch{1.5}
	\begin{tabular}{lcc} 
		\hline
		Line ratio &Best fit value&$3\sigma$ uncertainty\\
		\hline
$\log_{10}$([N\,{\sc ii}] / H$\alpha$) & -1.32 & 0.11 \\
$\log_{10}$([S\,{\sc ii}] / H$\alpha$)& -9.90 &0.06 \\
$\log_{10}$([O\,{\sc i}] / H$\alpha$)& -1.14 & 0.01 \\
$\log_{10}$([O\,{\sc iii}] / H$\beta$)&-1.20 & 0.07 \\
		\hline
	\end{tabular}
        \egroup
\end{table*}

\begin{figure*}
\begin{subfigure}[t]{0.35\textwidth}
    \includegraphics[width=\textwidth]{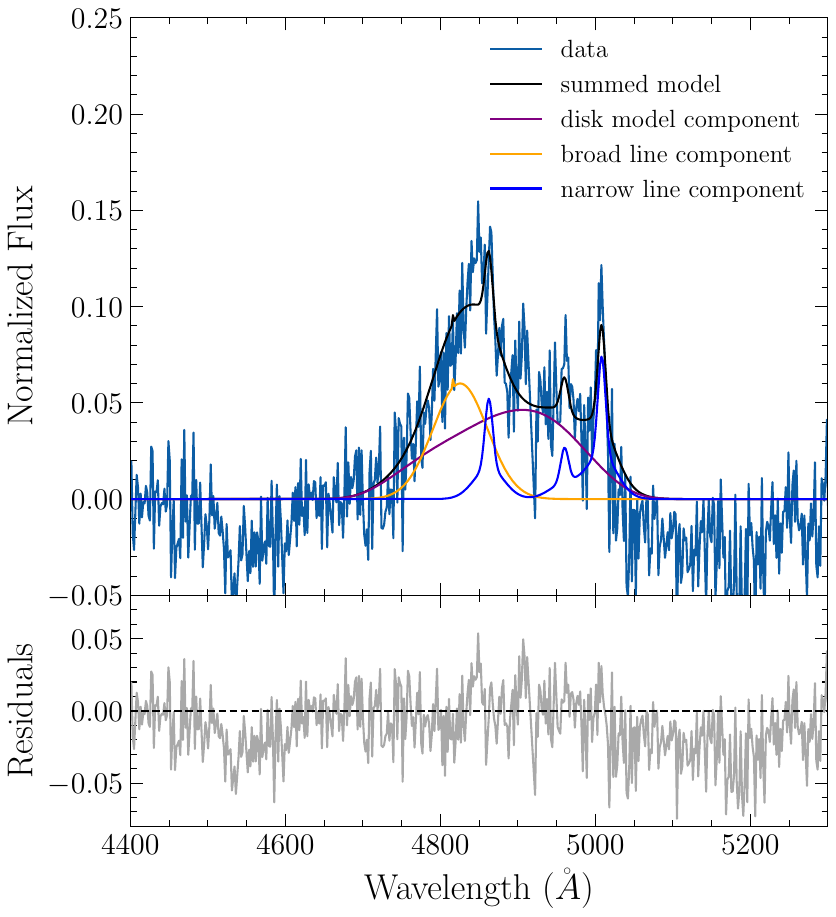}
    \caption{2017-03-24}
\end{subfigure}
\begin{subfigure}[t]{0.35\textwidth}
    \includegraphics[width=\linewidth]{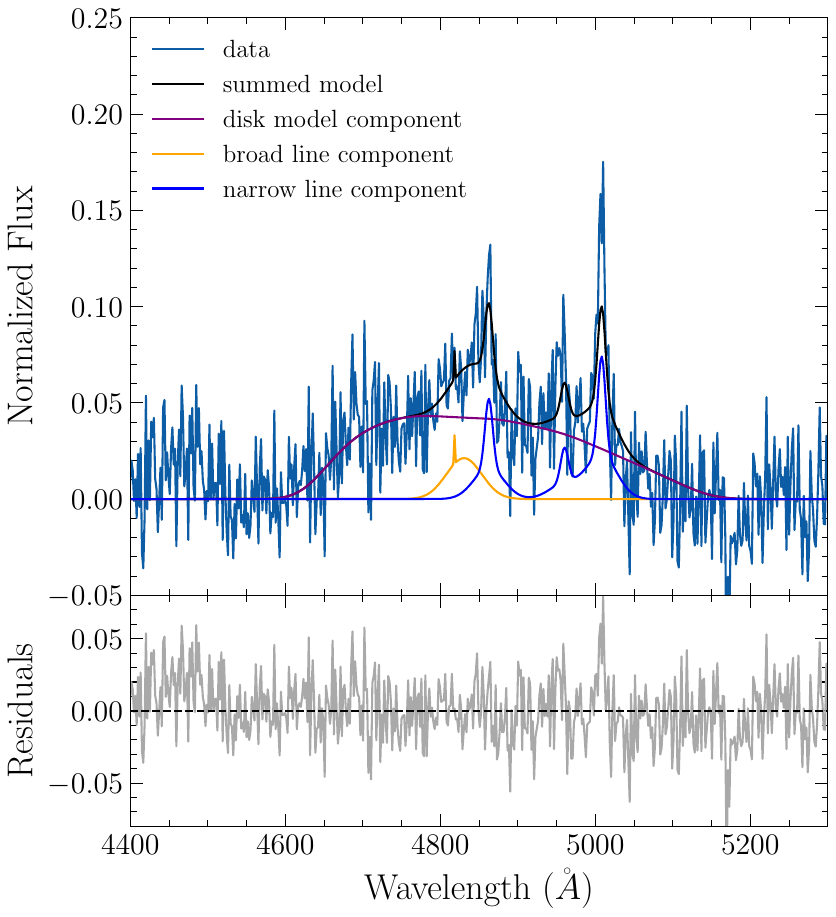}
    \caption{2018-01-05}
\end{subfigure}
\begin{subfigure}[t]{0.35\textwidth}
    \includegraphics[width=\linewidth]{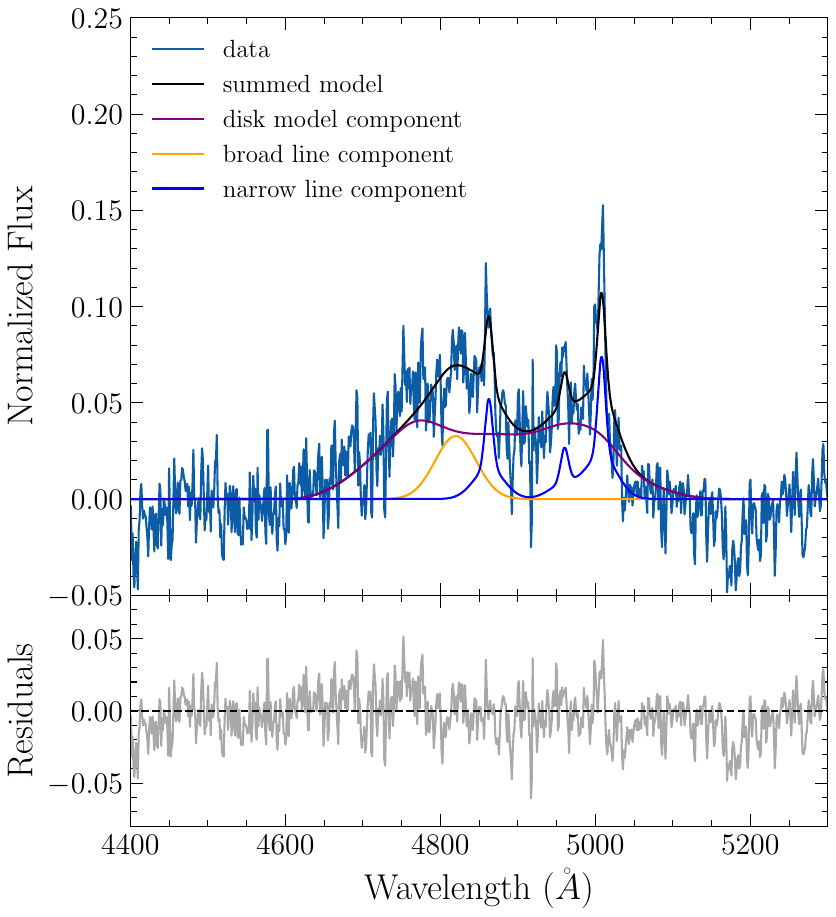}
    \caption{2018-05-15}
\end{subfigure}
\begin{subfigure}[t]{0.35\textwidth}
    \includegraphics[width=\linewidth]{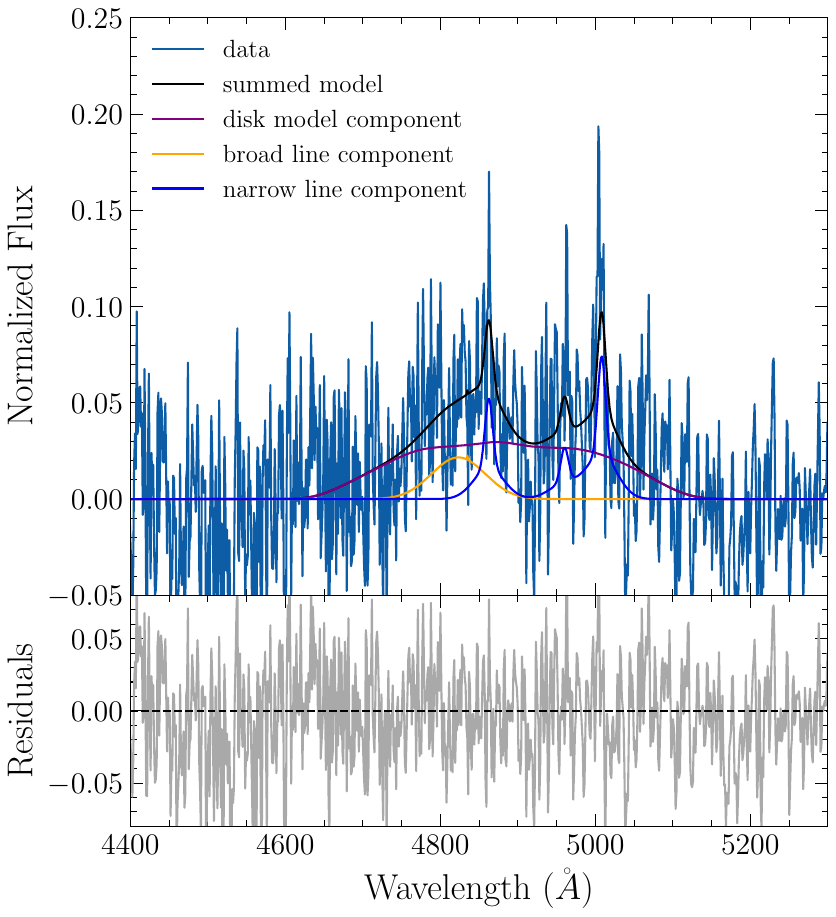}
    \caption{2019-01-26}
\end{subfigure}
\begin{subfigure}[t]{0.35\textwidth}
    \includegraphics[width=\linewidth]{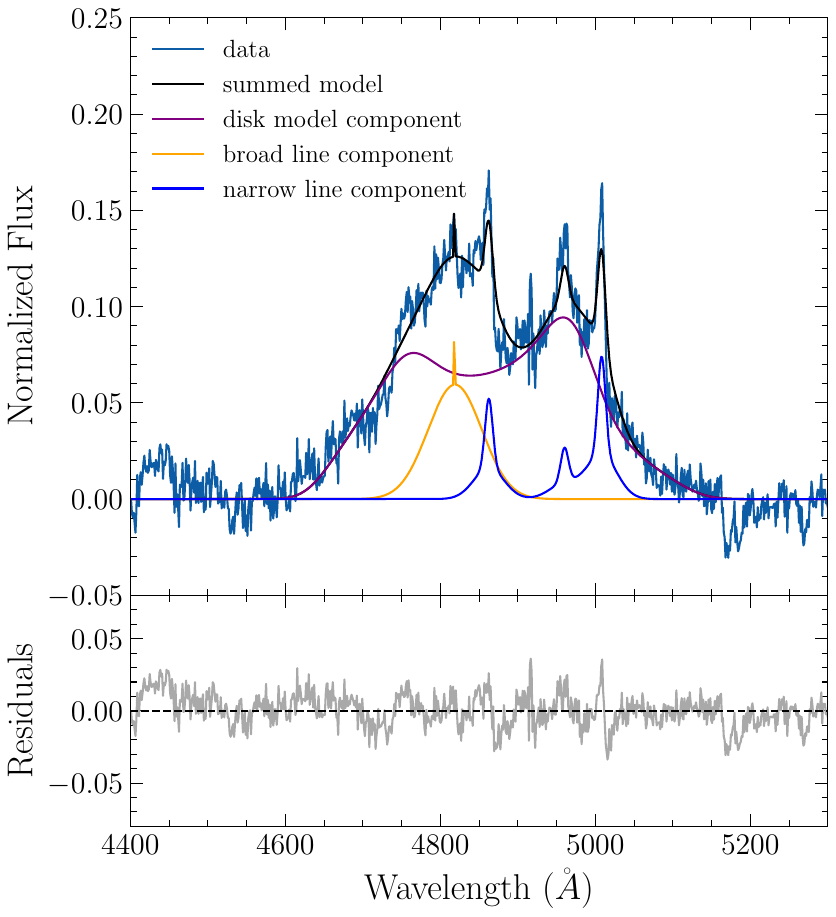}
    \caption{2020-01-28}
\end{subfigure}
\begin{subfigure}[t]{0.35\textwidth}
    \includegraphics[width=\linewidth]{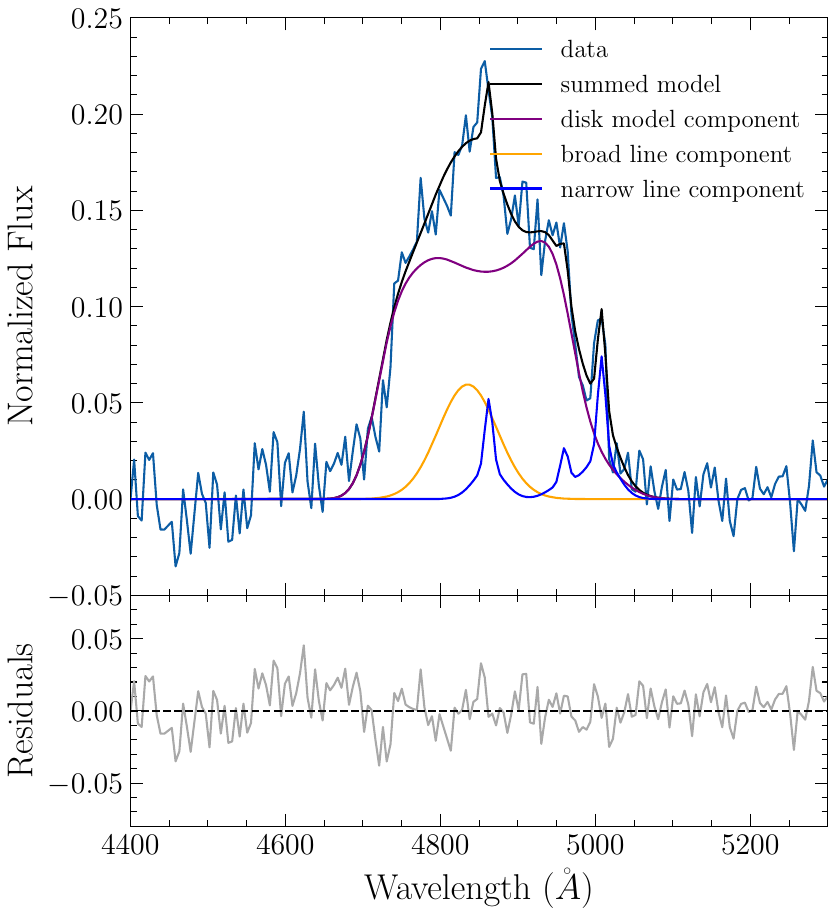}
    \caption{2023-01-15}
\end{subfigure}
\caption{Examples of the best-fit H$\beta$ disc and broad line models for 6 different spectral epochs of AT2017bcc corresponding to the fits shown in Figure \ref{fig:specfits}.}
\label{fig:specfitsbeta}
\end{figure*} 

\begin{figure*}
    \centering
    \includegraphics[width=\linewidth]{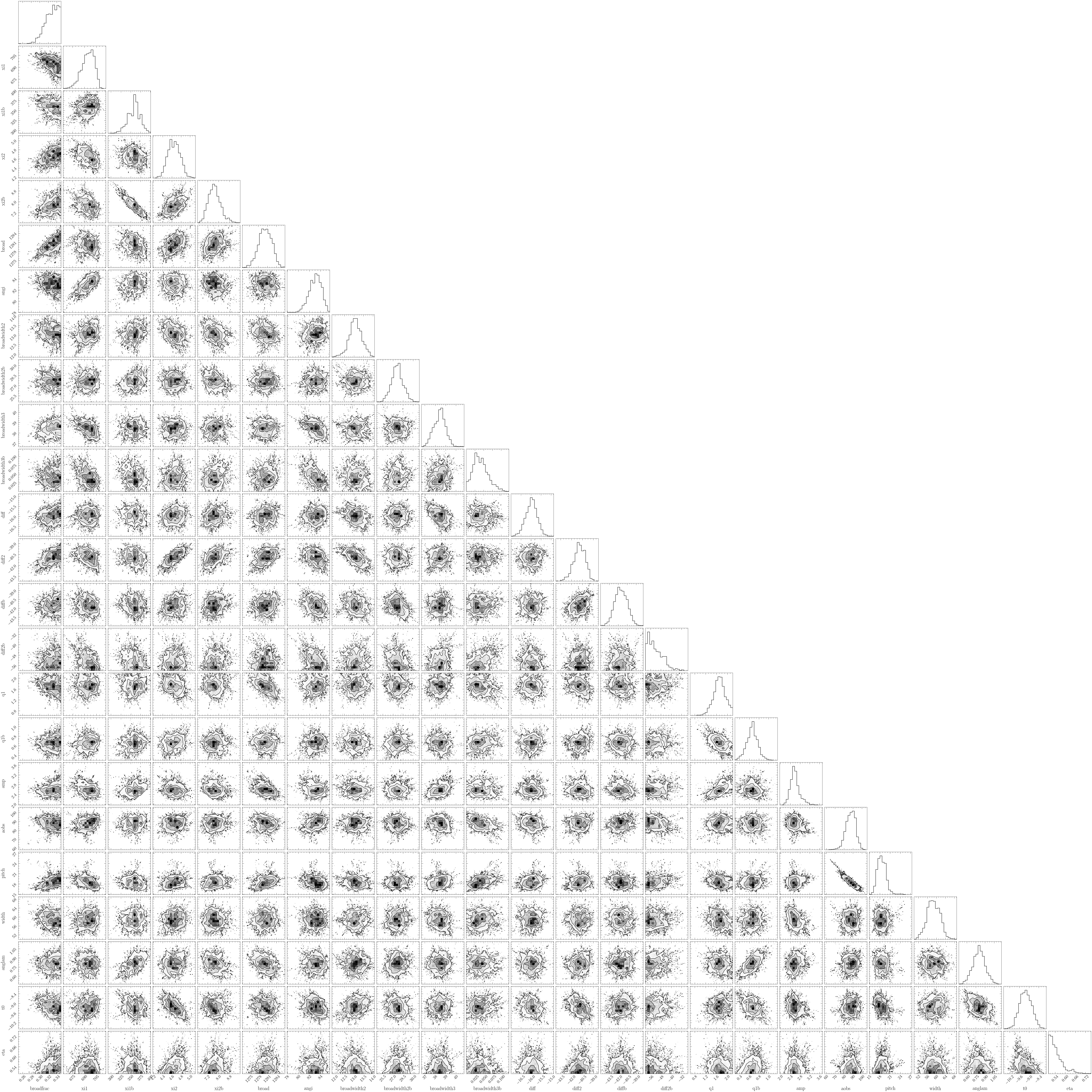}
    \caption{Corner plot for the parameters fit to the 2020-01-28 spectrum describing the disc and outflow components. From left to right we show the following parameters: the ratio of the amplitudes of the H$\beta$ disc profile and the H$\alpha$ disc profile, the inner radius of the H$\alpha$ emitting region of the disc $\xi_1$, the inner radius of the H$\beta$ emitting region of the disc $\xi_{1b}$, the outer radius of the H$\alpha$ emitting region of the disc $\xi_2$, the outer radius of the H$\alpha$ emitting region of the disc $\xi_{2b}$,  local turbulent broadening $\sigma$\ , disc inclination angle $i$, the widths of the two shark fin Gaussian components $\sigma_{b1}$ and $\sigma_{b2}$ and their rest wavelength offset from the narrow H$\alpha$ or H$\beta$ line central wavelengths $\Delta$ and $\Delta_b$, the emissivity power law indices for the H$\alpha$ emitting region and H$\beta$ emitting regions $q$ and $q_b$, the spiral arm amplitude $A_{\rm s}$, orientation angle $\phi$, pitch angle $\psi$ and width $w$, and the wind opening angle $\theta$, optical depth $\tau$, and the optical depth normalisation $t_0$.}
\end{figure*}

\begin{figure*}
    \centering
    \includegraphics[width=\linewidth]{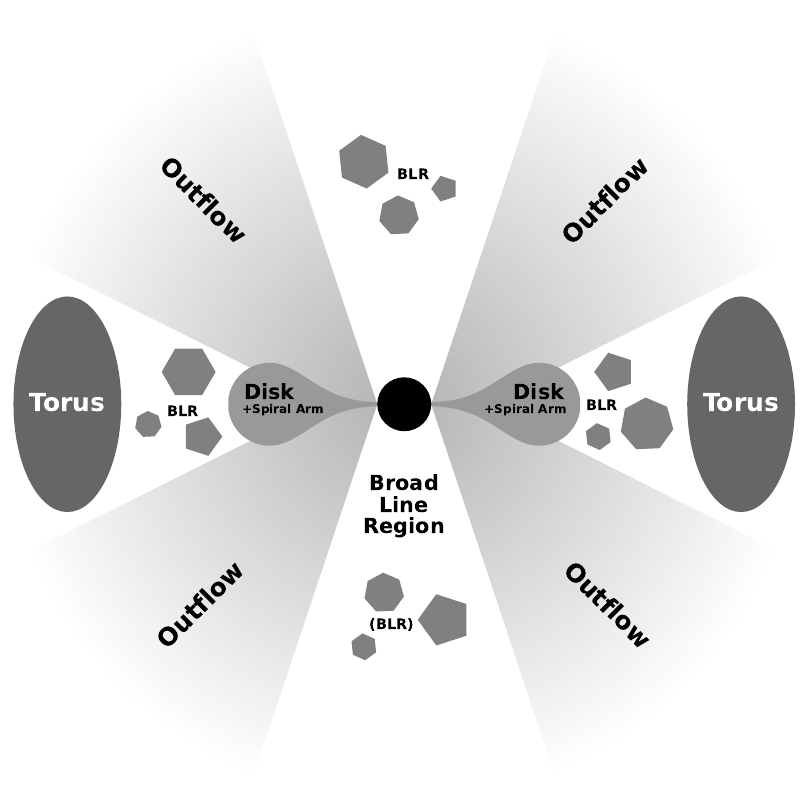}
    \caption{Cross-section diagram of the proposed emitting regions of AT2017bcc described in Section \ref{sec:profile-fitting}. The accretion disc with a spiral arm is shown around the central SMBH, and accounts for the double-peaked component of the emission profile. The double-Gaussian component is comprised of the broad line region (central Gaussian) and the outflow (blueshifted Gaussian). The narrow line emitting regions are expected to exist $\sim100$ light years from the galactic center, so are not shown here. Finally, the IR emission likely arises from a dusty torus surrounding the central SMBH, as it reprocesses emission from the AGN.}
\end{figure*}

\bsp	
\label{lastpage}
\end{document}